\documentclass[11pt]{article}
\pdfoutput=1
\usepackage{dcolumn}
\usepackage{bm}
\usepackage{graphicx}
\usepackage{amssymb,amsmath}
\usepackage{multirow}
\usepackage{cite,color,url}
\usepackage[colorlinks=true
,urlcolor=blue
,anchorcolor=blue
,citecolor=blue
,filecolor=blue
,linkcolor=blue
,menucolor=blue
,linktocpage=true
,pdfproducer=medialab
,pdfa=true3
]{hyperref}
\usepackage{authblk}

\usepackage{multicol}

\usepackage{slashed}
\usepackage{epsfig,psfrag,rotating,soul}
\usepackage{rotfloat}
\usepackage[font={small}]{caption}
\usepackage{colortbl}
\usepackage{xcolor}
\definecolor{darkspringgreen}{rgb}{0.09, 0.45, 0.27}

\evensidemargin \oddsidemargin
\title{Probing electroweak pair production of heavy neutral leptons \\ with displaced vertices at the LHC}

\author[a]{St{\'e}phane Lavignac\thanks{stephane.lavignac@ipht.fr}}
\author[b]{Anibal~D.~Medina\thanks{anibal.medina@fisica.unlp.edu.ar}}
\author[b]{Nicol\'as~I.~Mileo\thanks{mileo@fisica.unlp.edu.ar}}
\author[b]{Santiago Tanco\thanks{santiago.tanco@fisica.unlp.edu.ar}}

\affil[a]{\sl Institut de Physique Th{\'e}orique, Universit{\'e} Paris Saclay, CNRS, CEA, F-91191 Gif-sur-Yvette, France}
\affil[b]{\sl IFLP, CONICET - Dpto. \!de F\'{\i}sica, Universidad Nacional de La Plata, C.C. 67, 1900 La Plata, Argentina}

\date{}

\begin{document}
\maketitle

\begin{abstract}
    We study the sensitivity of displaced vertex searches at the LHC to heavy neutral leptons
(also known as sterile neutrinos) that are produced in pairs with an electroweak-size cross section.
We work within the context of a supersymmetric model in which the sterile neutrino
is produced along with Standard Model particles in higgsino decays.
By making use of model-independent reconstruction efficiencies provided by the ATLAS collaboration
in their search for displaced vertices with multiple jets, we obtain constraints on this model
from $139$ fb$^{-1}$ of data collected by ATLAS during the LHC Run~2,
and assess the discovery reach of Run~3 and of the high-luminosity LHC (HL-LHC).
Depending on the higgsino mass parameter, sterile neutrino masses between
$20~\mathrm{GeV}$ and $230~\mathrm{GeV}$ and active-sterile neutrino mixings in the range
$4 \times 10^{-14} \lesssim V^2_N \lesssim 3 \times 10^{-10}$ can be excluded. 
At the HL-LHC, discovery-level significances could be reached
for sterile neutrinos masses up to $295~\mathrm{GeV}$ and values of $V^2_N$ down to $3 \times 10^{-14}$.
Finally, moving away from the supersymmetric scenario, we study
to which extent these results can be generalized to a broader class of models
in which the sterile neutrinos are produced in the decays of heavier particles
that are themselves pair-produced with an electroweak-size cross section.
\end{abstract}

\flushbottom

\section{Introduction}

A large class of models of neutrino mass generation, mainly consisting of the type-I seesaw
mechanism~\cite{Minkowski:1977sc,Gell-Mann:1979vob,Yanagida:1979as,Glashow:1979nm,Mohapatra:1979ia} 
and of its variants, feature heavy sterile neutrinos. While these new states are often assumed to lie well above
the electroweak scale, like in Grand Unified realizations of the type-I seesaw mechanism,
there is no model-independent prediction for their masses.
In particular, they could live around the electroweak scale -- a possibility that may be natural in some
low-scale models such as the inverse seesaw mechanism~\cite{Wyler:1982dd,Mohapatra:1986aw,Mohapatra:1986bd} --
and be observable at
colliders~\cite{delAguila:2007qnc,Atre:2009rg,Deppisch:2015qwa,Banerjee:2015gca,Cai:2017mow,Mekala:2022cmm}.
Sterile neutrinos in this mass range are often referred to as heavy neutral leptons (HNLs) in the literature
(for a recent review, see e.g. Ref.~\cite{Abdullahi:2022jlv} and references therein).
In the simplest HNL model, the sterile neutrino $N$ only interacts via its mixing with
the Standard Model (SM) neutrinos (also known as active neutrinos), parameterized by the entries
$V_{N \alpha}$ ($\alpha = e, \mu, \tau$) of the $4 \times 4$ Pontecorvo-Maki-Nakagawa-Sakata  (PMNS) lepton mixing matrix.
At hadron colliders, the most studied production channel is $pp\to W^\pm \to l^{\pm} N$, followed
by $N \to l'^\pm l''^\mp \nu$ (trileptons) or $N \to l'^{\pm'} + 2~\mathrm{jets}$
(dileptons, which can be same sign if $N$ is a Majorana fermion).
In this process, both the sterile neutrino production and its decay are suppressed by the active-sterile neutrino mixing.
The $V_{N\alpha}$'s can be probed by studying the production of a serile neutrino along with
a charged lepton of flavour $\alpha$, but the sensitivity is limited by the fact that the production cross section
is proportional to $|V_{N\alpha}|^2$. Displaced vertex searches can probe smaller mixing angles thanks
to the highly suppressed SM backgrounds in the signal region (see e.g.
Refs.~\cite{Helo:2013esa,Antusch:2016vyf,Antusch:2017hhu,Cottin:2018nms,Abada:2018sfh,Drewes:2019fou,Cottin:2018kmq,Liu:2019ayx,Alimena:2019zri,DeVries:2020jbs,Cottin:2021lzz,Beltran:2021hpq,Urquia-Calderon:2023dkf,Wang:2024ieo,Bi:2024pkk}),
but the sensitivity remains limited. According to Ref.~\cite{Drewes:2019fou}, even using displaced vertices,
the high-luminosity  LHC (HL-LHC) will only be able to probe $m_N \lesssim 40~\mathrm{GeV}$
and $|V_{N e}|^2,  |V_{N \mu}|^2 \gtrsim 5 \times 10^{-10}$, well above the value suggested by he naive
seesaw formula $|V_{N \alpha}|^2 \sim m_\nu / m_N \sim 5 \times 10^{-12}\, (10\, \mathrm{GeV} / m_N)$.

In this paper, we consider an alternative scenario in which the sterile neutrino is produced
from the decays of heavier, beyond the Standard Model (BSM) particles $\psi$:
$\psi\to N+\mathrm{SM}$, with the sterile neutrino subsequently decaying into SM particles
via its mixing with active neutrinos, as in the standard case\footnote{An alternative possibility
is to have a pair of sterile neutrinos produced in the decays of a heavy BSM particle, such as a $Z'$
boson (see e.g. Refs.~\cite{Deppisch:2019kvs,Chiang:2019ajm,A:2025ygb}).
Due to the different topology, the experimental signatures differ from the ones of the scenario considered in this paper.}.
Assuming a production cross section of typical electroweak size for $\psi$,
the sterile neutrino production is not suppressed by the active-sterile neutrino mixing,
at variance with the standard HNL scenario.
Furthermore, should it be possible to measure the different $N$ decay channels and to reconstruct its total decay width,
one could access the active-sterile neutrino mixing angles~\cite{Lavignac:2020yld}.
This is easier to realize if the sterile neutrino has displaced decays.

An explicit realization of this non-standard scenario has been proposed in Ref.~\cite{Lavignac:2020yld}.
It is a supersymmetric model with $R$-parity violation in which the sterile neutrino is the supersymmetric
partner of a pseudo-Nambu-Goldstone boson and mixes with the higgsinos. As a consequence of this
mixing, the higgsinos decay to a sterile neutrino and SM particles with an almost $100\%$ branching fraction.
This leads to the production of pairs of sterile neutrinos in proton-proton collisions with an electroweak-size
cross section, numerically equal to the higgsino pair production cross section. Another feature of this
model is that two of the three SM neutrinos become massive at tree level, and that consistency with
neutrino oscillation data can be achieved by fixing some of the model parameters. The active-sterile neutrino
mixing is of order $\sqrt{m_\nu / m_N} \sim (10^{-7} - 10^{-6})$, resulting in displaced vertices from
sterile neutrino decays.

The goal of this paper is to study the sensitivity of displaced vertex searches
to sterile neutrinos that are pair-produced at the LHC with an electroweak-size cross section,
taking the model of Ref.~\cite{Lavignac:2020yld} as an example.
By recasting the ATLAS search for displaced vertices with multiple jets of Ref.~\cite{ATLAS:2023oti}
to this model, we obtain constraints on the sterile neutrino mass and on its mixing with active neutrinos
from the LHC Run~2 (corresponding to $\mathcal{L}=139$ fb$^{-1}$ of data collected by the ATLAS detector
at a center-of-mass energy $\sqrt{s}=13~\mathrm{TeV}$).
Note that many phenomenological studies based on displaced vertex searches are either model dependent
or assume an $100\%$ reconstruction efficiency, an assumpotion that is strongly violated at the LHC.
In order to assess the discovery reach of LHC Run~3 (which is taking data at a center-of-mass energy
$\sqrt{s}=13.6~\mathrm{TeV}$, and is expected to collect $\mathcal{L}=300$ fb$^{-1}$)
and of the HL-LHC ($\mathcal{L}=3$ ab$^{-1}$, $\sqrt{s}=14~\mathrm{TeV}$),
we make the well-motivated assumption of negligible SM background.
Finally, we investigate to which extent these results can be generalized to other, non-supersymmetric models
in which the sterile neurinos are produced in the decays of heavier particles with an electroweak-size
pair production cross section.

The paper is organized as follows. In Section~\ref{sec:model}, we present the sterile neutrino model
considered in this work. Section~\ref{sec:recasting} describes the procedure employed to recast the ATLAS search
for displaced vertices and multiple jets to the model of Section~\ref{sec:model}.
Section~\ref{sec:results} presents the constraints on the model obtained from this recasting,
as well as the discovery reach of LHC Run~3 and of the HL-LHC.
Finally, we give our conclusions in Section~\ref{sec:conclusion}.

%%%%%%%%%%%%%%%%%%%%%%%%%%%%%%
\section{The sterile neutrino model}
\label{sec:model}
%%%%%%%%%%%%%%%%%%%%%%%%%%%%%%

\begin{figure}
\centering
\includegraphics[width=0.4\linewidth]{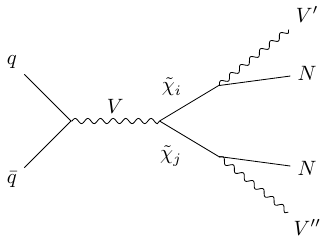}
\hskip 2.5cm
\includegraphics[width=0.26\linewidth]{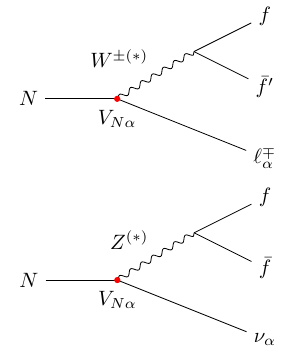} 
\vskip .1cm
\caption{Production (left) and decay (right) of the sterile neutrino. In the left diagram,
$V$, $V'$ and $V''$ are electroweak gauge bosons $W^\pm$ and $Z$, while $\tilde{\chi}_i$ and $\tilde{\chi}_j$
stand for higgsino-like electroweakinos $\tilde \chi^0_{1,2}$ and $\tilde \chi^\pm_1$.
In the right diagrams, the intermediate gauge bosons can be on shell or off shell, depending
on the sterile neutrino mass. The coefficients $V_{N \alpha}$ at the sterile neutrino--gauge boson
vertices are the active-sterile mixing angles, with $\alpha = e, \mu, \tau$.}
\label{fig:diagrams}
\end{figure}

The model considered in this paper is a particular realization of the scenario in which sterile neutrinos
are produced in the decays of heavy particles $\psi$, which are themselves pair-produced with an
electroweak-size cross section ($pp \to \psi \bar \psi$, $\psi \to N +$~SM particles).
It is a supersymmetric model with $R$-parity violation, in which the sterile neutrino is the supersymmetric
partner of a pseudo-Nambu-Goldstone boson and mixes with the higgsinos.
All supersymmetric partners are assumed to be decoupled, except for the electroweakinos.
We assume $M_1, M_2 \gg \mu > M_Z$, such that the lightest neutralinos
and chargino are almost pure higgsinos with nearly degenerate masses,
$m_{\tilde \chi^0_2} \simeq m_{\tilde \chi^\pm_1} \simeq m_{\tilde \chi^0_1} \simeq \mu$,
while $\tilde \chi^0_3$, $\tilde \chi^0_4$ and $\tilde \chi^\pm_2$ are gaugino-like and significantly
heavier.
The higgsino-like fermions are pair-produced in proton-proton collisions at the LHC and decay
promptly to a sterile neutrino and SM particles with an almost $100\%$ branching fraction.
The dominant decay modes are\footnote{The subdominant decays are
$\tilde \chi^\pm_1 \to N + X^\pm$ and $\tilde \chi^0_{1,2} \to N + X^0$,
where $X^\pm = \bar t b / t \bar b, Z W^\pm, t \bar t W^\pm, W^+ W^- W^\pm, \dots$ and
$X^0 = W^+ W^-, W^+ \bar t b, W^- t \bar b, \dots$. In all cases, two sterile neutrinos are produced
in each event.} $\tilde \chi^\pm_1 \to W^\pm + N$ and $\tilde \chi^0_{1,2} \to Z + N$,
leading to the production processes $p p \to \tilde \chi^0_1 \tilde \chi^0_2 \to Z Z N N$,
$p p \to \tilde \chi^\pm_1 \tilde \chi^0_{1,2} \to W^\pm Z N N$
and $p p \to \tilde \chi^+_1 \tilde \chi^-_1  \to W^+ W^- N N$.
The sterile neutrinos subsequently decay via on-shell or off-shell gauge bosons to the final states
$\ell^+ \ell^{\prime -} \nu$, $\ell^\pm q \bar q'$, $\nu q \bar q$, and $\nu \nu \nu$.
Their decay width is suppressed by the square of their
mixing angles with the active neutrinos, which are typically of order $\sqrt{m_\nu / m_N} \sim (10^{-7} - 10^{-6})$,
resulting in displaced vertices\footnote{The smallness of the active-sterile neutrino mixing angles 
also explains why the standard sterile neutrino production channels
$pp \to W^{\pm *} \to \ell^\pm_\alpha N$ and $pp \to Z^* \to \nu_\alpha N$ are negligible.}.
The sterile neutrino pair production and decays are depicted by the diagrams of Figure~\ref{fig:diagrams}.

%

%%%%%%%%%%%%%%%%%%%

Let us describe the main aspects of the model (we refer the reader to Ref.~\cite{Lavignac:2020yld}
for a more detailed discussion). We consider a supersymmetric extension of the SM with a global $U(1)$ symmetry
under which only the charged lepton and the down-type Higgs doublet superfields $L_i$ ($i = 1,2,3$)
and $H_d$ are charged.
Below the scale $f \gg M_W$ at which the global $U(1)$ symmetry is spontaneously broken,
the superpotential reads (omitting Yukawa couplings and trilinear $R$-parity violating couplings)
\begin{eqnarray}
  W\! & =\! & \mu_0\, H_u H_d + \mu_i\, H_u L_i + \lambda_0\, H_u H_d \Phi + \lambda_i\, H_u L_i \Phi
    +\, \dots\ ,
\label{eq:superpotential}
\end{eqnarray}
where $\Phi$ is a singlet chiral superfield containing the pseudo-Nambu-Goldstone boson $a$,
its CP-even scalar partner $s$ and their supersymmetric partner $\chi$.
Due to the superpotential terms $\lambda_i\, H_u L_i \Phi$, $\chi$ mixes with the SM neutrinos
and is therefore a sterile neutrino (it was dubbed pseudo-Goldstone sterile neutrino in Ref.~\cite{Lavignac:2020yld}).
Its mass $m_\chi$ arises predominantly from supersymmetry breaking\footnote{The same
is true for the CP-even scalar field $s$, whose mass is expected to be of the same
order as sfermions masses, while the pseudo-Nambu-Goldstone boson mass is solely due
to the explicit breaking of the global $U(1)$ symmetry, assumed to be small. The mass hierarchy
is therefore $m_a \ll m_\chi \ll m_s$, with $m_a \gtrsim 400\, \mbox{MeV}$ required
in order to evade cosmological and astrophysical bounds~\cite{Lavignac:2020yld}.}
and receives an irreducible contribution proportional to the gravitino mass~\cite{Cheung:2011mg}.
In this paper, we treat $m_\chi$ as a free parameter taking values between ${\cal O} (10)$ and
a few $100\, \mbox{GeV}$ (as we will see later, this is the range of values to which
the LHC is sensitive), with $m_\chi < \mu_0$ in order for the sterile neutrino to be produced
in higgsino decays.
The global $U(1)$ symmetry ensures that the $R$-parity violating parameters $\mu_i$ and $\lambda_i$
are suppressed relative to the corresponding $R$-parity conserving parameters $\mu_0$ and $\lambda_0$:
\begin{equation}
  \frac{\mu_i}{\mu_0}\ \sim\ \frac{\lambda_i}{\lambda_0}\ \sim\ \epsilon^{l-h_d}\, \ll\, 1\, ,
\label{eq:RPV_suppression}
\end{equation}
where $\epsilon$ is a small symmetry breaking parameter, and $l$ and $h_d$ (with $l > h_d > 0$)
are the $U(1)$ charges of the lepton and down-type Higgs doublets, respectively\footnote{ The same factor
$\epsilon^{l-h_d}$ supresses the trilinear $R$-parity violating couplings $\lambda_{ijk}$ and $\lambda'_{ijk}$
with respect to the charged lepton and down-type quark Yukawa couplings, such that the experimental upper bounds
on $R$-parity violating parameters (see e.g. Ref.~\cite{Barbier:2004ez} for a review) are easily evaded.}.
With the choice $h_d = 1$ made in Ref.~\cite{Lavignac:2020yld}, we also have $\lambda_0 \sim \mu_0 / f \ll 1$.
Note that the $U(1)$ symmetry is flavour universal, hence cannot explain the fermion mass hierarchy.

%%%%%%%%%%%%%%%%%%%

The superpotential terms~\eqref{eq:superpotential} induce a mixing between the higgsinos,
the SM neutrinos and the sterile neutrino $\chi$, making the neutralino mass matrix $M_N$
an $8 \times 8$ matrix. Similarly, the charged leptons mix with the charged higgsino,
resulting in a $5 \times 5$ chargino mass matrix $M_C$.
These mixings are suppressed by the small parameter ratios $\mu_i / \mu_0$ and $\lambda_i / \lambda_0$,
as well as by $\lambda_0$.
The three lightest neutralino mass eigenstates correspond to the light, mostly active neutrinos $\nu_i$
(where the indices $i = 1,2,3$ should not be confused with the generation indices appearing
in Eqs.~\eqref{eq:superpotential} and~\eqref{eq:RPV_suppression}),
while the fourth neutralino eigenstate, denoted by $N$,  is the mostly sterile neutrino with mass $m_N \simeq m_\chi$.
In the chargino sector, the three lightest states correspond to the physical charged leptons,
with masses $m_\alpha$ ($\alpha = e, \mu, \tau$).
The other neutralinos and charginos are mainly admixtures of higgsinos and gauginos
and are denoted by the usual MSSM notations.
Due to the assumption $M_1, M_2 \gg \mu_0 \gg M_Z$, $\tilde \chi^0_1$,  $\tilde \chi^0_2$
and $\tilde \chi^\pm_1$ are higgsino-like and nearly degenerate in mass, while 
$\tilde \chi^0_3$,  $\tilde \chi^0_4$ and $\tilde \chi^\pm_2$ are gaugino-like and significantly heavier.
In summary, the mass spectrum of the model is
\begin{eqnarray}
  & m_1,\, m_2,\, m_3\, \ll\, m_N\, \ll\, m_{\tilde \chi^0_1} \simeq m_{\tilde \chi^0_2}
  \ll m_{\tilde \chi^0_3} < m_{\tilde \chi^0_4}\,, \nonumber\\  &
  m_e \ll m_\mu \ll m_\tau \ll\, m_{\tilde \chi^\pm_1} \ll m_{\tilde \chi^\pm_2}\, ,
\label{eq:ino_spectrum}
\end{eqnarray}
where $m_N \simeq m_\chi$ and $m_{\tilde \chi^0_1} \simeq m_{\tilde \chi^0_2} \simeq m_{\tilde \chi^\pm_1} \simeq \mu$
(with $\mu \equiv \sqrt{|\mu_0|^2 + \sum_i |\mu_i|^2} \simeq |\mu_0|$).
The neutralinos and charginos mix among themselves, i.e. have off-diagonal couplings
to the $Z$ boson and non-standard couplings to the $W$ boson, leading to new interactions
that are absent from the SM. These include the following interactions of the sterile neutrino,
responsible for its production in higgsino decays and for its decays to SM particles:
\begin{equation}
  Z \tilde \chi^0_{1,2} N\, ,  \quad  W^\pm \tilde \chi^\mp_1 N\, ,  \quad
  Z N \nu_i\, ,  \quad  W^\pm N l^\mp_\alpha\, .
\end{equation}
In order to compute the higgsino and sterile neutrino decay rates, we
diagonalize the full $8 \times 8$ neutralino and $5 \times 5$ chargino mass matrices, and generalize
the standard expressions~\cite{Haber:1984rc} for the couplings $Z \tilde{\chi}^0_i \tilde{\chi}^0_j$,
$Z \tilde{\chi}^{\pm}_i \tilde{\chi}^{\mp}_j$ and $W^{\mp} \tilde{\chi}^{\pm}_i \tilde{\chi}^0_j$
to our model (with $i, j = 1 \dots 8$ for $\tilde{\chi}^0_{i, j}$ and $i, j = 1 \dots 5$ for $\tilde{\chi}^{\pm}_{i, j}$).

%%%%%%%%%%%%%%%%%%%

The number of free parameters in the model is considerably reduced by requiring consistency
with neutrino oscillation data. Due to the mixing between the SM neutrinos, the sterile neutrino
and the higgsinos, two of the light neutrino states become massive
($\nu_2$ and $\nu_3$ in the case of normal ordering, which we assume in this paper).
Taking advantage of the seesaw-like hierarchical structure of $M_N$, we perform an approximate
block-diagonalization to obtain an effective mass matrix for the light neutrinos:
\begin{equation}
  (M_\nu)_{\alpha\beta}\, \simeq\,  - \frac{\lambda^2_0 v^2_u}{m_\chi}
    \left( \frac{\mu_\alpha}{\mu_0} - \frac{\lambda_\alpha}{\lambda_0} \right)
    \left( \frac{\mu_\beta}{\mu_0} - \frac{\lambda_\beta}{\lambda_0} \right)
    - \frac{M_{11} M^2_Z \cos^2 \beta}{M_1 M_2}\, \frac{\mu_\alpha}{\mu_0}\, \frac{\mu_\beta}{\mu_0} \ ,
\label{eq:effective_Mnu}
\end{equation}
where $M_{11} \equiv \cos^2 \theta_W M_1 + \sin^2 \theta_W M_2$,
$v_u$ is the vacuum expectation value of $H^0_u$, and we are now using indices $\alpha = e, \mu, \tau$
to stress that we are working in the charged lepton mass eigenstate basis.
For fixed values of $\lambda_0$, $\mu_0$, $M_1$, $M_2$, $\tan \beta$ and $m_\chi$, the model parameters
$\mu_\alpha$ and $\lambda_\alpha$ can be expressed in terms of the light neutrino masses
$m_i$ (with $m_1 = 0$ following from the rank-2 structure of $M_\nu$ and from the assumption
of normal ordering), the PMNS matrix entries $U_{\alpha i}$,
and a complex number $z$ parametrizing a $2 \times 2$ complex orthogonal matrix $R$,
analogous to the Casas-Ibarra matrix of the seesaw mechanism~\cite{Casas:2001sr}:
\begin{equation}
    R\, =\, \begin{pmatrix}
    \cos{z} & \sin{z} \\
    - \sin{z} & \cos{z} 
    \end{pmatrix} .
\label{eq:R_matrix}
\end{equation}
Namely, one has
\begin{eqnarray}
  \frac{\mu_\alpha}{\mu_0} & = & \sqrt{\frac{M_1 M_2}{M_{11} M^2_Z \cos^2 \beta}}\,
    \left( R_{21} \sqrt{m_3}\, U^*_{\alpha 3} + R_{22} \sqrt{m_2}\, U^*_{\alpha 2} \right) , \\
  \frac{\lambda_\alpha}{\lambda_0} & = & \frac{\mu_\alpha}{\mu_0} - \frac{\sqrt{m_\chi}}{\lambda_0 v_u}
    \left( R_{11} \sqrt{m_3}\, U^*_{\alpha 3} + R_{12} \sqrt{m_2}\, U^*_{\alpha 2} \right) . 
\label{eq:mu_alpha_lambda_alpha}
\end{eqnarray}
Fixing the neutrino oscillation parameters $\Delta m^2_{31}$, $\Delta m^2_{21}$ and $\theta_{ij}$
to their best fit values~\cite{Esteban:2024eli} and neglecting CP violation in the PMNS matrix,
we are left with only 7 free parameters\footnote{We fixed the value of $\lambda_0$, which is related
to the global symmetry breaking scale $f$ by $\lambda_0 \simeq h_d \mu_0 / f$, by requiring
$\sum_\alpha |\mu_\alpha / \mu_0 - \lambda_\alpha / \lambda_0|^2 = \sum_\alpha |\mu_\alpha|^2 / \mu^2_0$.
This assumption has little impact on the sterile neutrino phenomenology, and could be relaxed
by treating $f$ as a free parameter of the model. Also, the matrix $R$ has a discrete ambiguity,
which we fixed by choosing $\det R = +1$ in Eq.~\eqref{eq:R_matrix}.}: $\mu$, $M_1$,
$M_2$, $\tan \beta$, $m_N$ and the complex mixing angle $z$. The model is therefore
rather predictive, and fully agrees with neutrino oscillation data as long as $\mbox{Im}\, z$
is not too large. Indeed, a large value of $\mbox{Im}\, z$ enhances $\mu_\alpha$ and $\lambda_\alpha$
by a factor $e^{|{\rm Im} z|}$, resulting in a fine-tuning in the light neutrino mass matrix~\eqref{eq:effective_Mnu}.
In this case, the seesaw approximation leading to Eq.~\eqref{eq:effective_Mnu} is no longer accurate.
In practice, we will restrict to $|{\rm Im}\, z| \leq 3$, which corresponds to the absence of a fine-tuning stronger
than $\approx 1\%$ in the light neutrino mass matrix\footnote{ 
The amount of fine-tuning in %the light neutrino mass matrix 
$M_\nu$ can be quantified by $f \equiv \mathrm{min}_{\{ \alpha, \beta \}}\, |(M_\nu)_{\alpha \beta}| /\!
\left| \frac{M_{11} M^2_Z \cos^2 \beta}{M_1 M_2}\, \frac{\mu_\alpha}{\mu_0}\, \frac{\mu_\beta}{\mu_0} \right|$
$= \mathrm{min}_{\{ \alpha, \beta \}} |(M_\nu)_{\alpha \beta}| / |(R_{21} \sqrt{m_3}\, U^*_{\alpha 3} + R_{22} \sqrt{m_2}\, U^*_{\alpha 2})
(R_{21} \sqrt{m_3}\, U^*_{\beta 3} + R_{22} \sqrt{m_2}\, U^*_{\beta 2})|$.
For large values of ${\rm Im}\, z$, $|R_{ij}| \simeq e^{|{\rm Im} z|} / 2$, hence $f \approx 4 e^{-2 |{\rm Im} z|}$.}.

%%%%%%%%%%%%%%%%%%%

The parameters $M_1$, $M_2$ and $\tan \beta$ have little impact on the collider signatures
of the model, as long as $M_1, M_2 > \mu$. They mainly affect the higgsino decay rates,
but the decays remain prompt and the  ``inclusive'' branching ratios ${\rm BR} (\tilde \chi^\pm_1 \to N + X)$
and ${\rm BR} (\tilde \chi^0_{1,2} \to N + X)$, where $X$ only contains SM particles,
are very close to $1$ in a broad region of the parameter space.
As a result, the sterile neutrino production cross section is to an excellent approximation
given by the higgsino pair production cross section
$\sigma (pp \to \tilde{\chi}^{\pm}_1 \tilde{\chi}^{\mp}_1, \tilde{\chi}^{\pm}_1 \tilde{\chi}^0_{1,2}, \tilde{\chi}^0_1 \tilde{\chi}^0_2)$.
The subsequent decays of the sterile neutrinos are computed using the couplings
between the electroweak gauge bosons and the neutralinos and charginos determined
from the diagonalization of $M_N$ and $M_C$. In practice, the interactions of $N$ with the $W$ and $Z$
bosons are very well approximated by the following effective Lagrangian, valid below the higgsino mass scale $\mu$:
\begin{equation}
  \frac{g}{2 c_W} \left( V_{N \alpha} Z_\mu \bar N \gamma^\mu \nu_\alpha + \mbox{h.c.} \right)
    + \frac{g}{\sqrt{2}} \left( V_{N \alpha} W^+_\mu \bar N \gamma^\mu \ell^-_\alpha + \mbox{h.c.} \right) ,
\label{eq:effective_N_interactions}
\end{equation}
where the $V_{N \alpha}$'s are the effective active-sterile neutrino mixing angles. They can be computed
by performing an approximate block-diagonalization of $M_N$ (analogous to the one leading
to Eq.~\eqref{eq:effective_Mnu}, but for the $4 \times 4$ effective neutrino mass matrix including
the sterile neutrino), yielding
\begin{equation}
  V_{N \alpha}\ \simeq\ R_{11}\, \sqrt{\frac{m_3}{m_N}}\, U^*_{\alpha 3}\,
    +\, R_{12}\, \sqrt{\frac{m_2}{m_N}}\, U^*_{\alpha 2}\, .
\label{eq:V_N_alpha}
\end{equation}
The overall active-sterile mixing is given by
\begin{equation}
  V_N\, \equiv\, \sqrt{\sum_\alpha |V_{N \alpha}|^2}\, .
\label{eq:V_N}
\end{equation}
We have checked that Eqs.~\eqref {eq:effective_N_interactions} and~\eqref{eq:V_N_alpha}
give a very good approximation to the true $W^\pm N l^\mp_\alpha$ and $Z N \nu_i$ couplings obtained
by diagonalizing exactly $M_N$ and $M_C$ (after summing over neutrino flavours in the case
of the $Z N \nu_\alpha$ couplings).
The scaling $V_N \sim \sqrt{m_\nu / m_N}$, reminiscent of the seesaw mechanism, results in
tiny mixing angles\footnote{In fact, the active-sterile mixing angles and the light neutrino masses
are controlled by the same ratios of parameters $\mu_i / \mu_0$ and $\lambda_i / \lambda_0$,
which are small by virtue of approximate $R$-parity conservation (the dependence is quadratic
for the $m_i$'s and linear for the $V_{N \alpha}$'s). Hence, approximate $R$-parity conservation
explains the smallness of both neutrino masses and active-sterile neutrino mixing.},
leading to displaced vertices from the sterile neutrino decays.

In addition to the sterile neutrino mass, the active-sterile mixing angles also depend
on the complex parameter $z$. One can single out two particular values of $z$,
corresponding to what we will call the {\it minimal mixing} ($z = \pi/2$) and
{\it maximal real mixing} ($z = 0$) cases. This terminology is justified by the fact that,
for real $z$, $V_N$ varies between $V^{\rm min}_N \equiv V_N (z = \pi/2) \simeq \sqrt{m_2 / m_N}$
and $V^{\rm max}_N \equiv V_N (z = 0) \simeq \sqrt{m_3 / m_N}$.
For complex $z$, $V_N$ can be larger than $V^{\rm max}_N$ (and even much larger,
since $V_N \propto e^{|{\rm Im} z|}$ for $|{\rm Im}\, z| > 1$), but it cannot be smaller than $V^{\rm min}_N$.

%%%%%%%%%%%%%%%%%%%%%%%%%%%%%%%%%%%
\section{Recasting the ATLAS multijet search with displaced vertices}
\label{sec:recasting}
%%%%%%%%%%%%%%%%%%%%%%%%%%%%%%%%%%%

Long-lived particles (LLPs) are intensely searched for at the LHC by the ATLAS, CMS and LHCb collaborations. A great variety of signatures are part of the search programme, including displaced vertices (DVs), displaced leptons, disappearing tracks, trackless jets, among others. The results are interpreted in terms of BSM models that predict LLPs, examples of which are SUSY models such as R-parity violating SUSY~\cite{ATLAS:2019fwx,ATLAS:2020xyo,ATLAS:2023oti,CMS:2020iwv,CMS:2021tkn,CMS:2021kdm}, gauge-mediated SUSY breaking~\cite{CMS:2021kdm,CMS:2020iwv,ATLAS:2020wjh}, stealth SUSY~\cite{ATLAS:2018tup} and split SUSY~\cite{CMS:2020iwv,ATLAS:2017tny,ATLAS:2018yey}; models with a hidden (dark) sector that communicates with the SM via a scalar portal~\cite{ATLAS:2018niw,CMS:2019ajt,ATLAS:2022gbw,ATLAS:2022izj,CMS:2022qej}; and models of neutrino mass generation~\cite{ATLAS:2022atq,CMS:2022nty}.
Although these analyses may in many cases be sensitive to a larger variety of models and, in particular, could provide constraints on the model considered in this paper, it is difficult to perform a thorough recasting without additional model-independent information on the detector response. We found three LLP analyses that provide the auxiliary material necessary for a proper recasting of the results, namely, searches for long-lived massive particles in events with displaced vertices along with missing transverse energy~\cite{ATLAS:2017tny}, oppositely charged leptons~\cite{ATLAS:2019fwx}, and multiple jets~\cite{ATLAS:2023oti}. In this work, we focus on the last one since
the model described in Section~\ref{sec:model} is characterized by a larger branching ratio into jets than into light charged leptons, due to the presence of massive gauge bosons in the decays of the electroweakinos and of the sterile neutrino.
Moreover, the large  multiplicity of jets with high transverse momentum ($p_T$) present in the signal makes the multijet trigger applied in Ref.~\cite{ATLAS:2023oti} more efficient, in contrast to models in which $N$ is produced through the active-sterile mixing angles.

The ATLAS DV search in the multijet channel~\cite{ATLAS:2023oti} targets signals from electroweakinos that decay into three quarks via a small $R$-parity violating coupling.
Two signal regions (SR) are used in the search: the High-$p_T$ SR and the Trackless jet SR.
The High-$p_T$ SR targets pair-produced gluinos that decay into a long-lived neutralino $\tilde \chi^0_1$ and a quark-antiquark pair.
The Trackless jet SR, on the other hand, targets electroweakino pair production signals. In both cases, two LLPs are produced.
This search provides model-independent reinterpretation material, in the form of event-level and vertex-level acceptance requirements and parameterized efficiencies that must be applied to truth-level simulation data.
Selected events pass jet multiplicity and $p_T$ thresholds, depending on the SR:
\begin{itemize}
    \item High-$p_T$ SR: at least 4, 5, 6 or 7 jets with $p_T>250,195,116$ or $90$~GeV, respectively.
    \item Trackless jet SR: at least 4, 5, 6 or 7 jets with $p_T>137,101,83$ or $55$~GeV, respectively, and at least 1 or 2 displaced jets with $p_T>70$ or $50$~GeV, respectively.
\end{itemize}
In the Trackless jet SR, a displaced jet is defined as a truth jet matched with the decay position of the LLP\footnote{A trackless jet in the ATLAS search is defined as a jet for which the scalar sum of the transverse momenta of all standard tracks is less than $5$~GeV~\cite{ATLAS:2023oti}. This requirement is replaced in the recasting material by the condition on displaced jets, which uses truth-level information.}.
The event-level efficiency is parameterized in terms of $\sum p_T^{\rm truth\,jet}$ for three different regions of $R^{\rm max}_{\rm decay}$, which is the largest (of the two produced LLPs) transverse distance between the interaction point and the LLP decay.
These efficiencies depend on the SR and are given in Figs.~1 and 2 of the auxiliary file~\cite{auxiliary-info}
associated with the ATLAS paper~\cite{ATLAS:2023oti}.

\vskip .1cm

The vertex-level acceptance requirements are:
\begin{itemize}
    \item The decay position must lie inside the fiducial volume ($R_{xy}<300~{\rm mm}$, $|z|<300~{\rm mm}$),
where $R_{xy}$ is the transverse distance from the interaction point, and $z$ the longitudinal coordinate.
    \item The transverse distance $R_{\rm decay}$ from the interaction point to the decay vertex must be larger than $4~{\rm mm}$.
    \item At least 1 charged particle from the truth vertex decay products must have a transverse impact parameter $|d_0|>2$~mm
(see definition below).
    \item The number of selected decay products (i.e., tracks associated with the decay vertex) must be at least 5. Here selected decay products refer to charged decay products that are stable for timescales required to traverse the tracking volume
and have an electric charge $q$ and a transverse momentum $p_T$ such that $p_T/|q| > 1$ GeV.
    \item The invariant mass of the truth decay vertex must be larger than 10 GeV. For consistency with the DV reconstruction in the experimental analysis, the truth decay vertex is constructed using the momenta of the selected decay products and assuming a charged pion mass.
\end{itemize}
The transverse impact parameter is defined as
the distance of closest approach of the particle track to the interaction point in the transverse plane,
and can be calculated as $|d_0|=R_{\rm decay}\sin(\Delta\phi)$, where $\Delta\phi$ is the azimuthal angle between the particle momentum and the vector from the interaction point to the decay vertex.
The definition of the various displaced observables can be found in Ref.~\cite{Allanach:2016pam}.
The vertex-level efficiencies are provided as functions of the
number of tracks associated with the DV, its invariant mass
and $R_{\rm decay}$ (all at truth level) in Figs.~3 and 4 of the auxiliary file~\cite{auxiliary-info},
and are independent of the SR.

In order to recast the ATLAS analysis~\cite{ATLAS:2023oti} to the model presented in Section~\ref{sec:model},
we write the model in \texttt{Feynrules 2.3}~\cite{Alloul:2013bka} and export it in \texttt{UFO 2.0} format~\cite{Degrande:2011ua}.
Then we generate parton-level event samples of $pp\to \tilde{\chi}_i \tilde{\chi}_j \to NNVV'$ with \texttt{MadGraph 3.5.2}~\cite{Alwall:2014hca},
where $\tilde{\chi}_i$, $\tilde{\chi}_j$ stand for higgsino-like electroweakinos $\tilde \chi^0_{1,2}$ and $\tilde \chi^\pm_1$,
and $V,V'$ are electroweak gauge bosons $W^\pm$ and $Z$. The calculation of displaced lengths during the event generation is enabled via the time-of-flight option.
The possible electroweakino production channels are $\tilde{\chi}^{\pm}_1 \tilde{\chi}_{1}^{0}, \tilde{\chi}^{+}_1 \tilde{\chi}^{-}_1, \tilde{\chi}_{1}^{0}\tilde{\chi}_{2}^{0}$ and $\tilde{\chi}^{\pm}_1 \tilde{\chi}_{2}^{0}$, due to their higgsino-like nature.
The cross sections for each production channel are calculated with \texttt{Resummino 3.1.2}~\cite{Fiaschi:2023tkq} at NLO+NLL precision.
We use \texttt{Pythia 8.3}~\cite{Bierlich:2022pfr} to simulate $N$ decays, showering and hadronization, and \texttt{FastJet 3.4}~\cite{Cacciari:2011ma} for jet clustering, using the anti-$k_T$ algorithm with $R=0.4$.
Both tools are used within our own code, adapted from Ref.~\cite{code-LLP-recasting}
to the model of Section~\ref{sec:model}.
We validated our method by reproducing the exclusion curve for one of the models considered
in the ATLAS search~\cite{ATLAS:2023oti} (see Appendix~\ref{sec:validation}).
We apply the event and vertex selection cuts and efficiencies to truth-level data, as required by the recasting material~\cite{auxiliary-info}.
No detector simulation is performed, as detector effects are contained in the event and vertex reconstruction efficiencies provided by the ATLAS collaboration.
Finally, we obtain event weights that are used to calculate signal yields for each point in the parameter space.
Note that the recasting is done for an integrated luminosity of $\mathcal{L}=139\;{\rm fb}^{-1}$, and our results are extrapolated to larger luminosities by assuming the same reconstruction efficiencies and zero background\footnote{No SM particles with high masses are expected to yield displaced decays satisfying all the selection requirements listed previously (at event- and vertex-level). As pointed out in~\cite{ATLAS:2023oti}, the search for multi-track DVs benefits from a small background arising mostly from instrumental and algorithmic effects. Note that the DV reconstruction efficiencies provided in~\cite{auxiliary-info} are binned by the transverse distance of the LLP decay in order to take into account the effect of the material map veto used in the full analysis to remove background DVs from hadronic interactions.}.

In our numerical simulations, we fix $M_1 = 3\, \mbox{TeV}$, $M_2 = 6\, \mbox{TeV}$ and $\tan \beta = 10$.
We vary the higgsino mass parameter $\mu$ between $500\, \mbox{GeV}$ and $2\, \mbox{TeV}$,
and the sterile neutrino mass $m_N$ between ${\cal O} (10)$ and a few $100\, \mbox{GeV}$.
The complex angle $z$ parametrizing the arbitrariness in the neutrino sector is also taken
as a free parameter, with the restriction $|{\rm Im}\, z| \leq 3$ in order to avoid fine-tuning
in the light neutrino mass matrix (see Section~\ref{sec:model} for details).
We do not consider values of $\mu$ lower than $500\, \mbox{GeV}$,
which could conflict with negative results from searches for neutralinos and charginos at the LHC.
While most of these experimental analyses are model dependent, one can recast some of them
to obtain constraints on $\mu$ in the model considered in this paper.
The most stringent constraint we found comes from a combination of several CMS searches
for electroweak production of wino-like charginos and neutralinos in the channel
$p p \to \tilde{\chi}^{\pm}_1 \tilde{\chi}^0_2\to W^{\pm} Z \tilde{\chi}^0_1 \tilde{\chi}^0_1$~\cite{CMS:2024gyw},
where $\tilde{\chi}^0_1$ is the (stable) lightest supersymmetric particle, assumed to be bino-like.
Recasting the CMS analysis for the process $\tilde{\chi}^{\pm}_1 \tilde{\chi}^0_2\to W^{\pm} Z N N$
(where $\tilde{\chi}^{\pm}_1$ and $\tilde{\chi}^0_2$ are higgsino-like and the sterile neutrinos decay
to three neutrinos, such that their decay products are not detected),
we obtain a lower bound on $\mu$ that strongly depends on the sterile neutrino mass\footnote{This is due
to the fact that the CMS upper bound on the $\tilde{\chi}^{\pm}_1 \tilde{\chi}^0_2$ production cross section,
which is essentially independent of $m_{\tilde \chi^0_1}$ for $m_{\tilde \chi^0_1} \lesssim 200\, \mbox{GeV}$,
is valid for the product $\sigma (pp \to {\tilde{\chi}^{\pm}_1 \tilde{\chi}^0_2}) \times [\mbox{BR} (N \to 3\nu)]^2$
in the sterile neutrino model, where the value of $\mbox{BR} (N \to 3\nu)$ strongly depends on $m_N$
-- especially on whether the sterile neutrino decays through on-shell or off-shell electroweak gauge bosons.} $m_N$.
For instance, we find $\mu \gtrsim 220\, \mbox{GeV}$ ($135\, \mbox{GeV}$) for $m_N = 70\, \mbox{GeV}$ ($110\, \mbox{GeV}$).
Taking into account the fact that some visible decays of the sterile neutrinos may be missed,
and allowing for an improvement of the CMS bound at Run~3, we conservatively assume
$\mu \geq 500\, \mbox{GeV}$ throughout this paper.

%%%%%%%%%%%%%%%%
\section{Run 2 constraints and projected reach of Run 3 and HL-LHC}
\label{sec:results}
%%%%%%%%%%%%%%%%

In this section, we present the results of the recasting of the ATLAS search for multijets with diplaced
vertices~\cite{ATLAS:2023oti} to the model described in Section~\ref{sec:model}, following the procedure
explained in Section~\ref{sec:recasting}. We obtain in this way the regions of the model parameter space
that are excluded by LHC Run~2. Then, extrapolating our results to higher energies and luminosities, we discuss
the expected discovery reach of Run~3 and of the high-luminosity LHC (HL-LHC).

\begin{table}[]
    \centering
    \begin{tabular}{c|c|c}
        & Trackless jet SR & High-$p_T$ SR \\ \hline
        Initial events & 5090 & 5090 \\
        Jet selection & $533$ & $51.1$ \\
        Decay position $\in$ fiducial volume & $88.1$ & $7.44$ \\
        $R_{\rm decay} > 4~{\rm mm}$ & $86.7$ & $7.44$ \\
        $\geq 1$ charged particle with $|d_0|>2\ {\rm mm}$ & $80.5$ & $6.85$ \\
        $\geq 5$ selected decay products & $66.8$ & $5.68$ \\
        $m_{\rm DV}>10~{\rm GeV}$ & $65.4$ & $5.48$ \\
        DV reconstruction efficiency & $3.58$ & $0.56$  
    \end{tabular}
    \caption{Cutflows for the model of Section~\ref{sec:model}, assuming $\mu=500~\mathrm{GeV}$, minimal mixing
and $m_N=70~\mathrm{GeV}$, based on the recasting of the ATLAS analysis~\cite{ATLAS:2023oti}.
The number of events corresponds to $\sqrt{s}=13~\mathrm{TeV}$ and $\mathcal{L}=139~\mathrm{fb}^{-1}$.}
    \label{tab:cuflow70}
\end{table}
\begin{table}[]
    \centering
    \begin{tabular}{c|c|c}
        & Trackless jet SR & High-$p_T$ SR \\ \hline
        Initial events & 5090 & 5090 \\
        Jet selection & $869$ & $69.1$ \\
        Decay position $\in$  fiducial volume & $869$ & $69.1$ \\
        $R_{\rm decay} > 4~{\rm mm}$ & $312$ & $30.7$ \\
        $\geq 1$ charged particle with $|d_0|>2~{\rm mm}$ & $253$ & $25.1$ \\
        $\geq 5$ selected decay products & $233$ & $22.5$ \\
        $m_{\rm DV}>10~{\rm GeV}$ & $233$ & $22.5$ \\
        DV reconstruction efficiency & $127$ & $14.6$ 
    \end{tabular}
    \caption{Same as Table~\ref{tab:cuflow70}, but for $m_N=150~\mathrm{GeV}$.}
    \label{tab:cutflow150}
\end{table}

Let us first consider the recasting of the ATLAS analysis, which used $\mathcal{L}=139~\mathrm{fb}^{-1}$
of $pp$ collision data collected at a center-of-mass energy $\sqrt{s}=13~\mathrm{TeV}$ during the LHC Run~2.
Tables~\ref{tab:cuflow70} and~\ref{tab:cutflow150} show the cutflows (number of events
passing the different cuts, selection criteria and efficiencies) for the model of Section~\ref{sec:model}
with two different sterile neutrino masses, $m_N = 70$~GeV (Table~\ref{tab:cuflow70}) and 
$m_N = 150$~GeV (Table~\ref{tab:cutflow150}). In both cases,
the higgsino mass parameter is fixed to $\mu = 500\, \mathrm{GeV}$,
the minimal mixing case (see definition at the end of Section~\ref{sec:model}) is assumed,
and the cutflows are shown for the Trackless jet and  High-$p_T$ signal regions.
One can readily see that the jet selection criteria eliminate about 10 times more events
in the High-$p_T$ SR than in the Trackless jet SR,
something expected given the relatively low value of $\mu$, which corresponds
to the mass of the neutralinos and charginos involved in the sterile neutrino production. 
The impact of the  fiducial volume cut strongly depends on the sterile neutrino mass:
while the number of events is reduced by a factor $\approx 6$ for $m_N = 70\, \mathrm{GeV}$,
it is left unchanged for $m_N = 150\, \mathrm{GeV}$.
This reflects the fact that the sterile neutrino decay length strongly varies with its mass
(with $c \tau_N \propto m^{-5}_N$ for $m_N < M_W$ and $c \tau_N \sim m^{-3}_N$ for $m_N > M_Z$).
For $m_N = 150\, \mathrm{GeV}$, practically all sterile neutrinos decay in the tracker,
while for $m_N = 70\, \mathrm{GeV}$, $c \tau_N$ is much larger and most decay products
land outside the tracking volume.
The dependence of the sterile neutrino decay length on its mass also explains why
the $R_{\rm decay} > 4~{\rm mm}$ cut has much more impact for $m_N = 150\, \mathrm{GeV}$
than for $m_N = 70\, \mathrm{GeV}$.

\begin{figure}[t]
    \centering
    \includegraphics[width=0.48\linewidth]{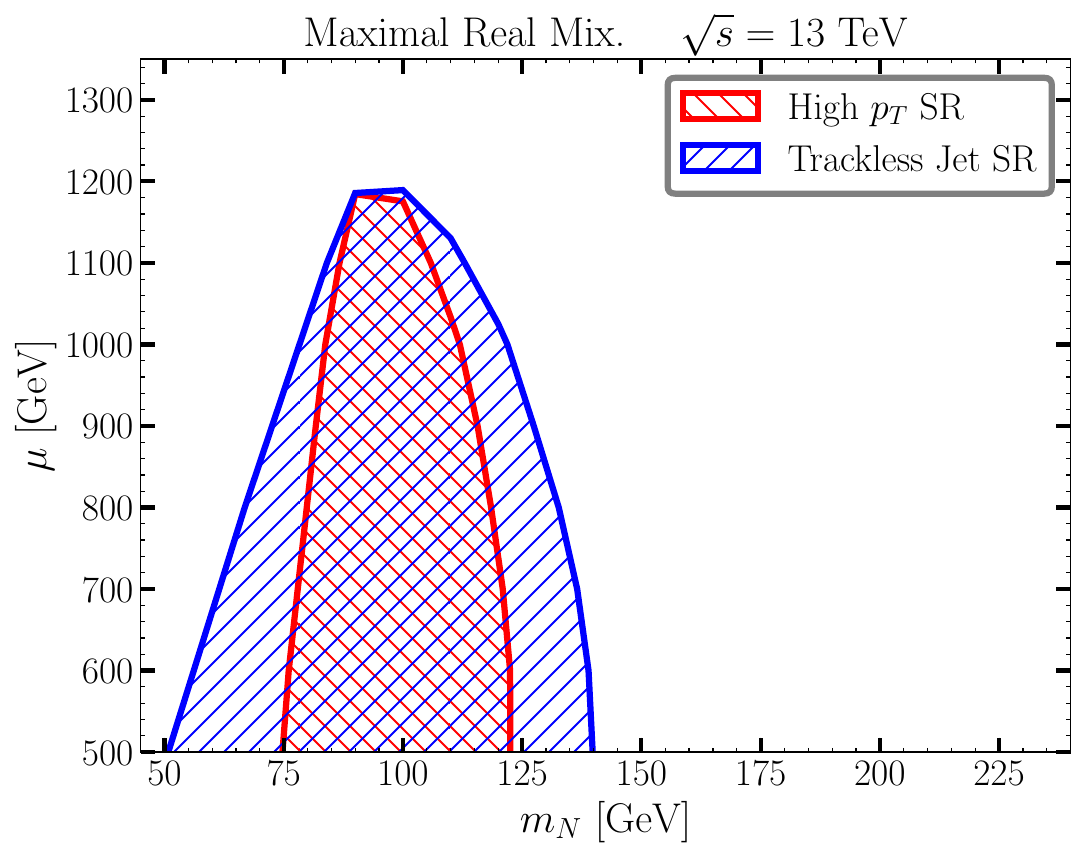}
    \includegraphics[width=0.48\linewidth]{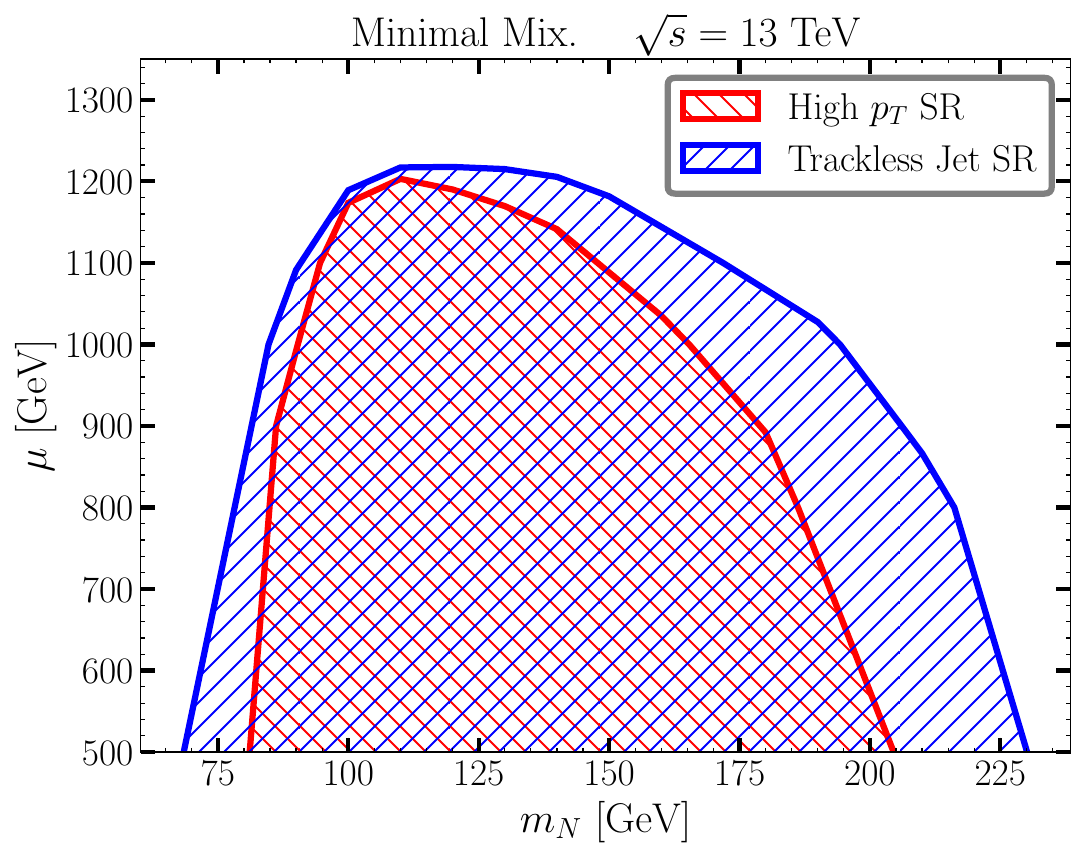}
    \caption{$95\%$ C.L. exclusion regions in the $(m_N, \mu)$ parameter space of the model of Section~\ref{sec:model}
from the recasting of the ATLAS analysis~\cite{ATLAS:2023oti},
assuming maximal real mixing (left plot) and minimal mixing (right plot).
}
    \label{fig:13000_exclusion}
\end{figure}

Figure~\ref{fig:13000_exclusion} shows the regions of the $(m_N, \mu)$ parameter space that are excluded
at the $95\%$ confidence level by the recasting of the ATLAS DV + multijet search, for two different mixing scenarios:
the maximal real mixing case (left plot) and the minimal mixing case (right plot).
The excluded area corresponding to the Trackless jet signal region is displayed in blue,
while the one associated with the High-$p_T$ signal region is in red.
The left border of each exclusion region is determined by the fiducial volume cut: since the sterile neutrino
decay length increases when $m_N$ decreases, most decays occur outside the tracking volume
for low $m_N$. As for the right border, it is controlled by the acceptance requirement $R_{\rm decay} > 4~{\rm mm}$,
as heavier sterile neutrinos tend to decay promptly.
The first thing one can notice in Figure~\ref{fig:13000_exclusion} is that the exclusion region extends
over larger sterile neutrino masses in the minimal mixing scenario than in the maximal real mixing case.
This is due to the fact that the sterile neutrino decay length is inversely proportional to $V^2_N$,
such that the fraction of prompt decays for a given $m_N$ is larger
in the maximal real mixing case than for minimal mixing. 
One can also see that the excluded area includes smaller $m_N$ values
in the maximal real mixing scenario, because the fraction of sterile neutrinos
that decay within the tracking volume is larger than in the minimal mixing case for such low values of $m_N$.
Another notable feature of Figure~\ref{fig:13000_exclusion} is that the excluded ranges of $m_N$ values
shrink as $\mu$ increases, as expected due to the decrease in the higgsino production cross section.
The maximal excluded value of $\mu$ lies around $1.2~\mathrm{TeV}$ for both signal regions.
Let us finally note that the High-$p_T$ excluded regions are completely covered by the Trackless jet ones.
This can be understood from the fact that the High-$p_T$ SR requires more energetic jets than the Trackless jet SR,
a criterion that is rather selective at the Run~2 energy and luminosity (see Tables~\ref{tab:cuflow70} and~\ref{tab:cutflow150}).

\begin{figure}[t]
    \centering
    \includegraphics[width=0.48\linewidth]{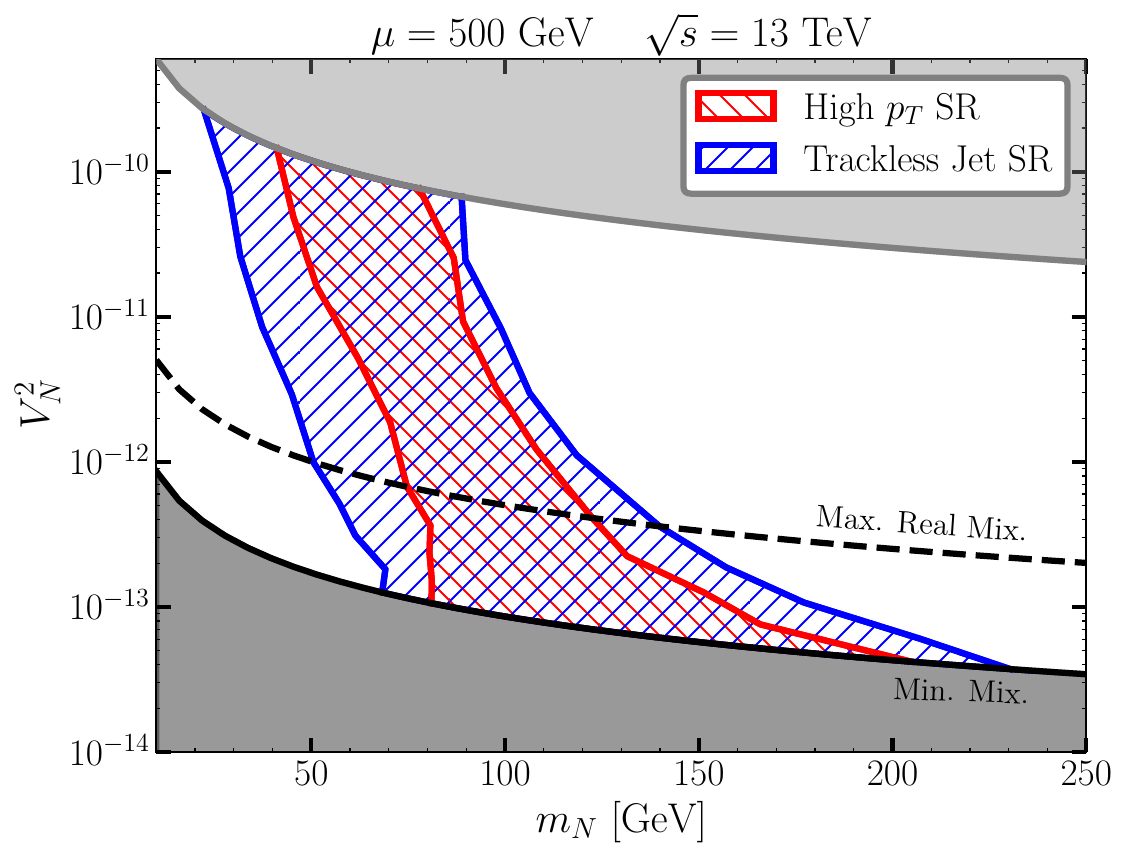}
    \includegraphics[width=0.48\linewidth]{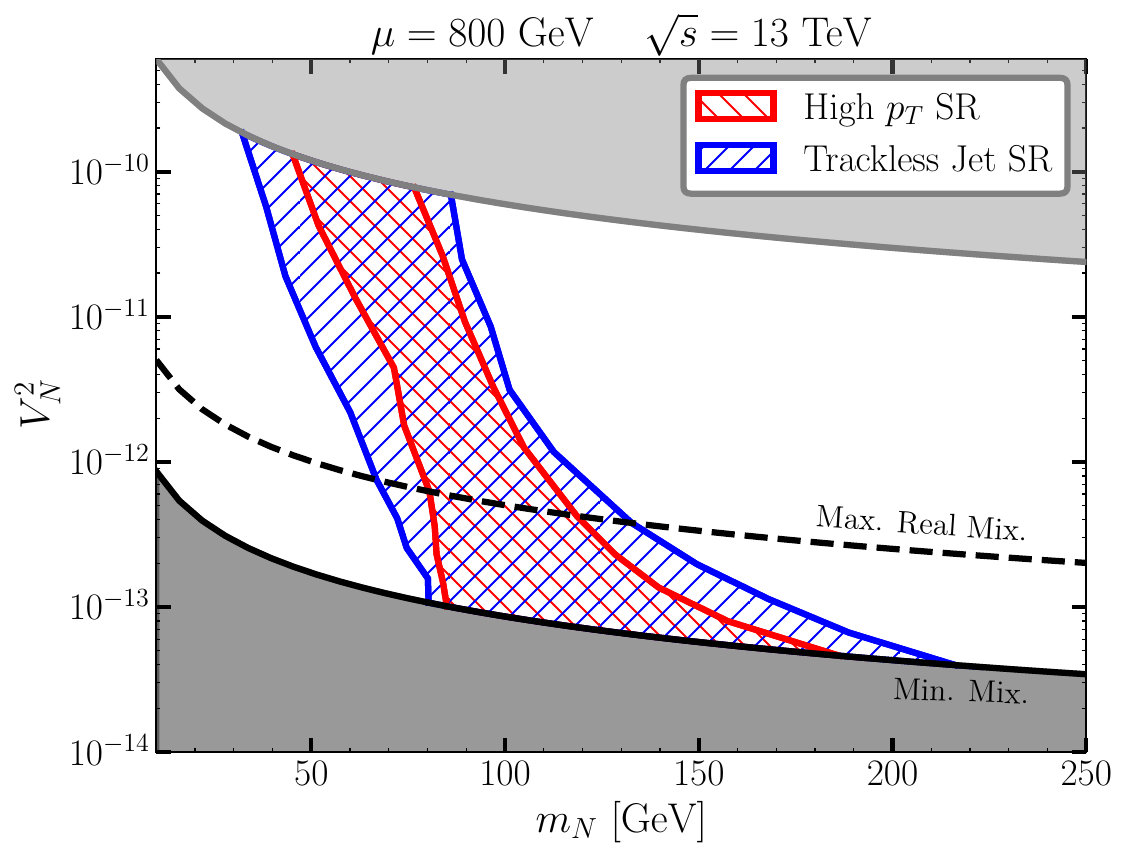}
    \caption{$95\%$ C.L. exclusion regions in the $(m_N, V^2_N)$ parameter space of the model of Section~\ref{sec:model}
from the recasting of the ATLAS analysis~\cite{ATLAS:2023oti},
assuming $\mu = 500~\mathrm{GeV}$ (left plot) and $\mu = 800~\mathrm{GeV}$ (right plot).
The bottom dark grey area is inconsistent with neutrino oscillation data, while the top light grey region
corresponds to a fine-tuning larger than $\approx 1\%$ in the light neutrino mass matrix.}
    \label{fig:13000_500}
\end{figure}

So far we only considered the minimal mixing and maximal real mixing cases. 
We now move on to other values of the active-sterile mixing, quantified
by\footnote{Stricly speaking, the active-sterile neutrino mixing is parametrized by the three quantities $V_{N \alpha}$
($\alpha = e,\, \mu,\, \tau$), which as shown by Eq.~\eqref{eq:V_N_alpha} depend on $m_N$ and on a complex parameter $z$.
For a given $m_N$, there is a continuous set of values of $z$ corresponding to the same $V_N$, but to different $V_{N \alpha}$'s. 
However, since the recasted ATLAS search is not sensitive to lepton flavour, the results presented in this section
depend mainly on $V_N$ and little on the actual $V_{N \alpha}$ values
(there is a small sensitivity to $V_{N \tau}$ coming from the decays $N \to \tau^\pm W^{\mp *}$ with the $\tau^\pm$
decaying hadronically).
This makes it possible to present approximate exclusion regions in the $(m_N, V^2_N)$ plane.}
$V_N \equiv \sqrt{\sum_\alpha |V_{N \alpha}|^2}$ ($\alpha = e,\, \mu,\, \tau$).
Figure~\ref{fig:13000_500} displays the exclusion areas
corresponding to the Trackless jet and High-$p_T$ SRs in the $(m_N, V^2_N)$ plane
for a fixed value of $\mu$ ($\mu = 500~\mathrm{GeV}$ and $800~\mathrm{GeV}$ for the left and right plots, respectively).
The black solid curve corresponds to the minimal mixing case, i.e. to the smallest value of $V_N$
consistent with the observed neutrino oscillation parameters, namely $V^{\rm min}_N = \sqrt{m_2 / m_N}$.
The bottom dark grey area is therefore inconsistent with experimental neutrino data.
The black dashed curve is associated with maximal real mixing,
defined as the maximal value that $V_N$ can reach for real values of the $z$ parameter,
namely $V^{\rm max}_N = \sqrt{m_3 / m_N}$.
Finally, the top border curve corresponds to $z=3i$, which as discussed in Section~\ref{sec:model}
implies a fine-tuning of about $1\%$ in the light neutrino mass matrix. The top light grey area
would lead to a larger fine tuning, and has been excluded from our analysis for that reason.
Note that the parameter space visible on Figure~\ref{fig:13000_500} is unconstrained by
neutrinoless double beta decay. Indeed, the latest lower limit on $T^{0 \nu}_{1/2}$ from
the KamLAND-Zen collaboration, $T^{0 \nu}_{1/2} (^{136}\mathrm{Xe}) \geq 3.8 \times 10^{26}\; \mathrm{yr}$
at 90\% C.L.~\cite{KamLAND-Zen:2024eml}, yields the upper bound (updated from Figure~3 of Ref.~\cite{Faessler:2014kka})
$|V_{Ne}|^2 \leq (1-2.5) \times 10^{-9}\,( m_N / 1\, \mathrm{GeV})$, valid for $m_N \gtrsim 1~\mathrm{GeV}$.

As in Figure~\ref{fig:13000_exclusion}, the left boundaries of the exclusion regions are set by the fiducial volume cut,
while the right borders are associated with the acceptance requirement $R_{\rm decay} > 4~{\rm mm}$.
The different shapes of these boundaries are due to the fact that for $m_N < M_W$, sterile neutrinos decay
via off-shell $W$ and $Z$ bosons, with $c \tau_N \propto m^{-5}_N V^{-2}_N$,
while for $m_N > M_Z$ the intermediate gauge bosons are on shell and $c \tau_N \sim m^{-3}_N V^{-2}_N$.
This explains the flatness of the right border curves at small $V^2_N$ and large $m_N$ values.
As can be seen from the plots, for $\mu \leq 800~\mathrm{GeV}$,
LHC Run~2 excludes values of the active-sterile neutrino mixing
in the range $4 \times 10^{-14} \leq V^2_N \leq 2 \times 10^{-10}$, depending on the sterile neutrino mass.
Note that Run~2 could also exclude larger values of $V^2_N$ that lie in the fine-tuned, light grey areas.

\begin{figure}[t]
    \centering
    \includegraphics[width=0.48\linewidth]{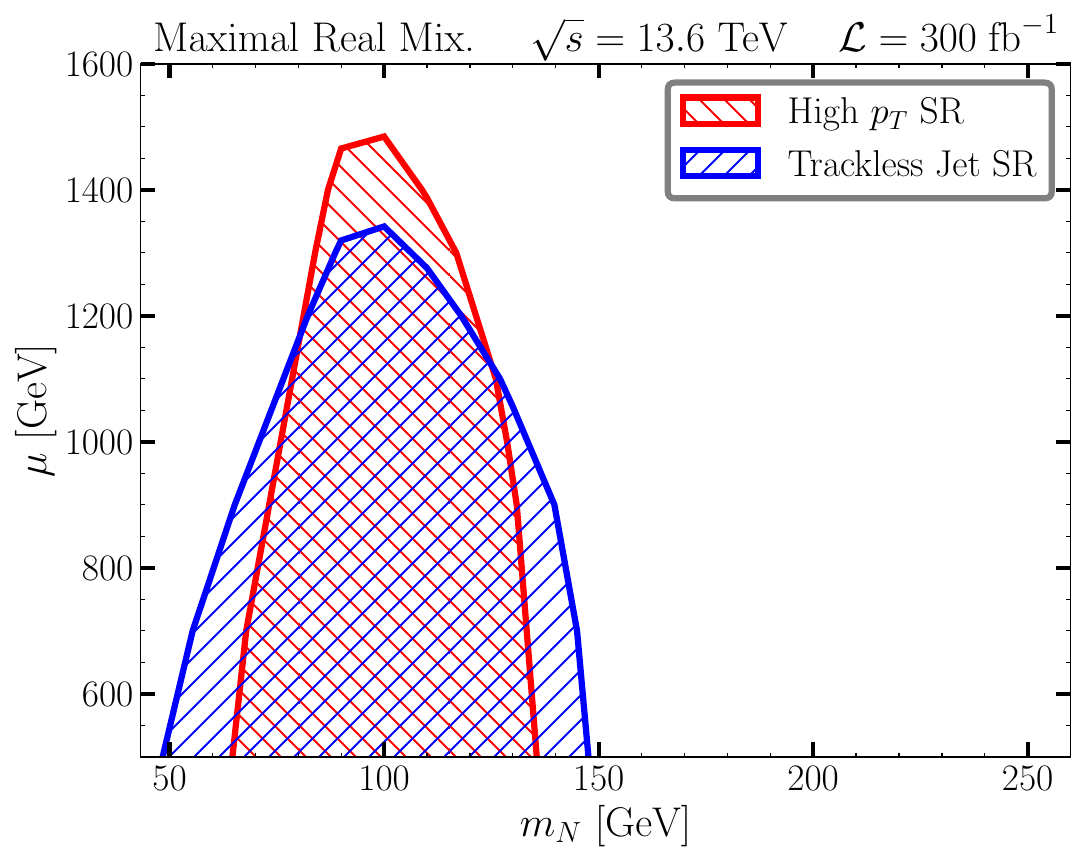}
    \includegraphics[width=0.48\linewidth]{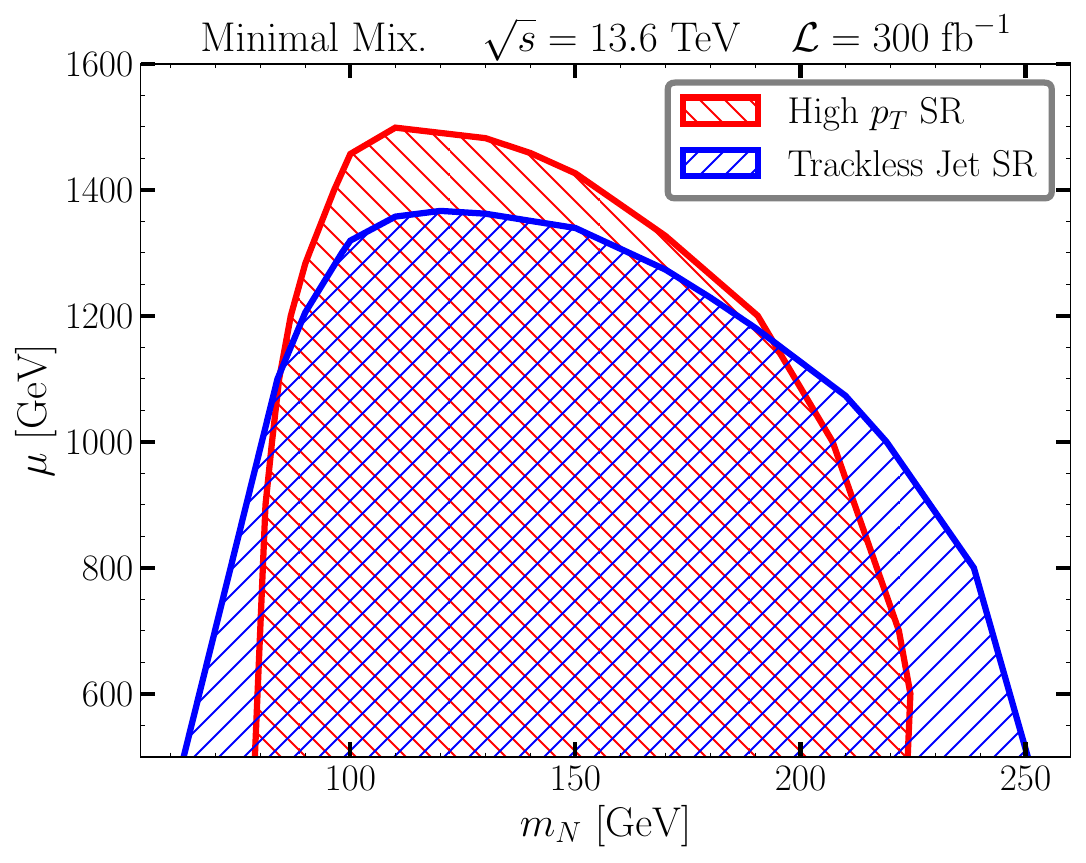}
    \caption{Projected discovery regions in the $(m_N, \mu)$ parameter space of the model of Section~\ref{sec:model}
for LHC Run~3 with $\mathcal{L}=300~\mathrm{fb}^{-1}$, assuming maximal real mixing (left plot) and minimal mixing (right plot).}
    \label{fig:13600_disc}
\end{figure}

\begin{figure}[t]
    \centering
    \includegraphics[width=0.48\linewidth]{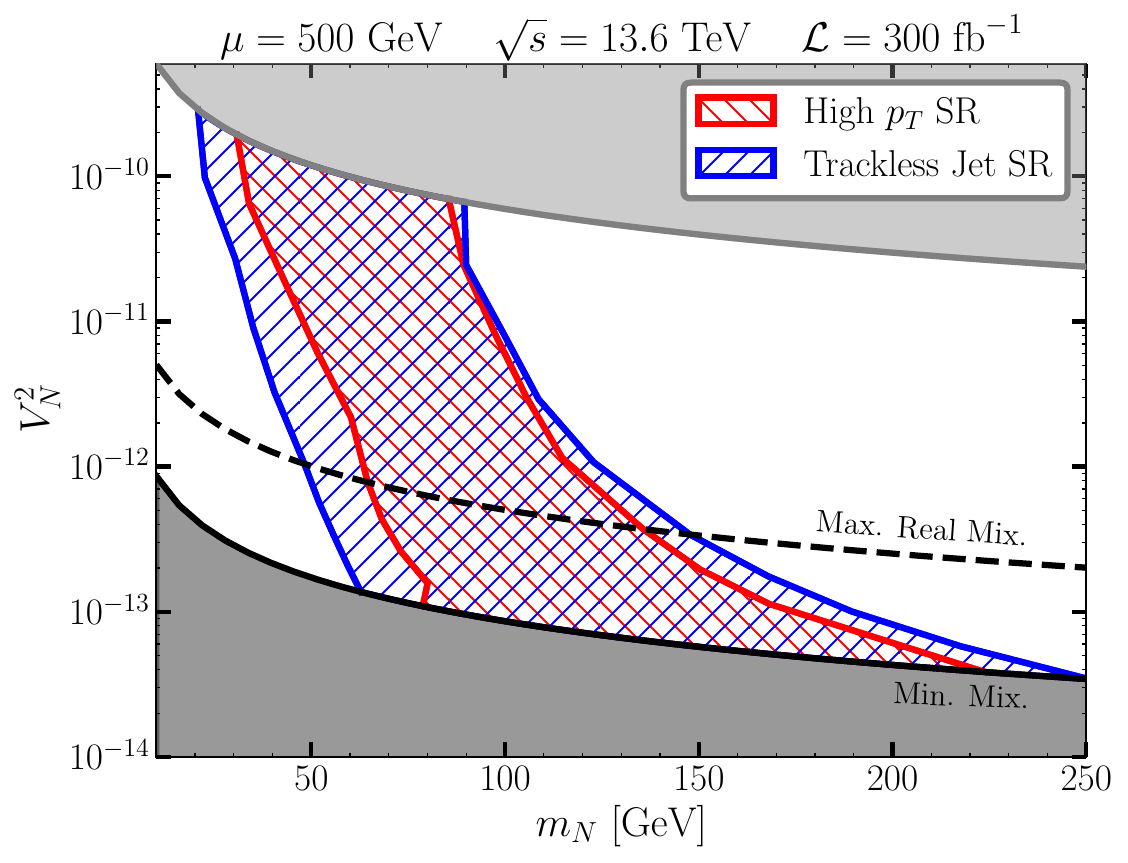}
    \includegraphics[width=0.48\linewidth]{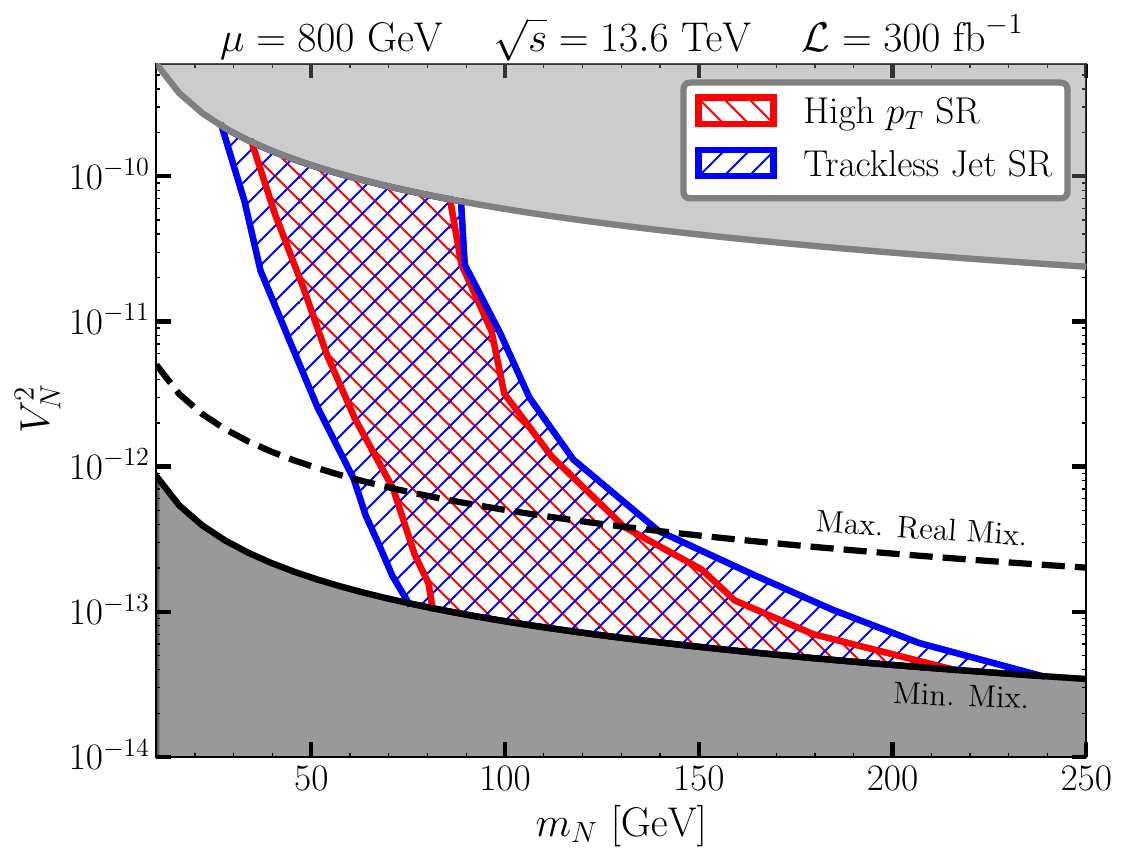} \\
    \includegraphics[width=0.48\linewidth]{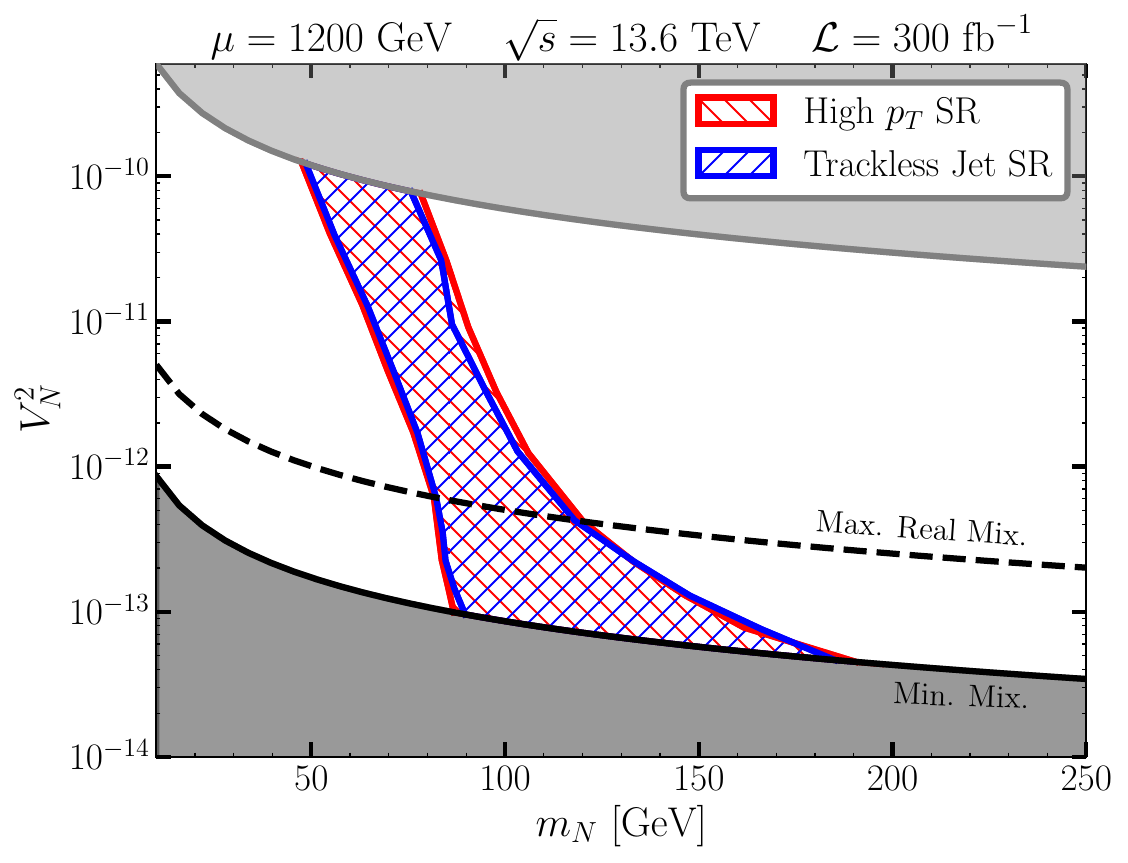}
    \caption{Projected discovery regions in the $(m_N, V^2_N)$ parameter space of the model of Section~\ref{sec:model}
for LHC Run~3 with $\mathcal{L}=300~\mathrm{fb}^{-1}$, assuming $\mu = 500~\mathrm{GeV}$ (upper left plot),
$800~\mathrm{GeV}$ (upper right plot) and $1.2~\mathrm{TeV}$ (lower plot).}
    \label{fig:13600_500}
\end{figure}

Having derived constraints on the model of Section~\ref{sec:model} by recasting
the ATLAS multijet + DV analysis of LHC Run~2 data~\cite{ATLAS:2023oti}, we can now extrapolate
the procedure to higher energies and luminosities in order to assess the discovery reach
of LHC Run~3 and of the HL-LHC.
Figure~\ref{fig:13600_disc} shows the projected discovery areas in the $(m_N, \mu)$ plane
for the center-of-mass energy of Run~3 ($\sqrt{s}=13.6~\mathrm{TeV}$) and an integrated luminosity
of $\mathcal{L}=300~\mathrm{fb}^{-1}$, assuming maximal real mixing (left plot) or minimal mixing (right plot).
As discussed in Section~\ref{sec:recasting}, we assumed zero background from SM processes,
such that discovery can be claimed with only 3 events.
Comparing Figure~\ref{fig:13600_disc} with Figure~\ref{fig:13000_exclusion}, one can see that
the larger energy and luminosity of Run~3 allow for a wider range of $m_N$ and $\mu$ values
to be probed in both mixing scenarios (up to  $\mu \approx 1.35~\mathrm{TeV}$ for the Trackless jet SR
and $\mu \approx 1.5~\mathrm{TeV}$ for the High-$p_T$ SR).
Notice that the High-$p_T$ SR now overtakes the Trackless jet SR for $\mu \gtrsim 1.15~\mathrm{TeV}$,
since large multiplicities of energetic jets are more easily produced
for such higgsino masses\footnote{Also, in the Trackless jet SR, the event-level efficiencies drop
at large jet transerve momenta, while they remain approximately constant  in the High-$p_T$ SR
(see Figures~1 and~2 of the auxiliary file~\cite{auxiliary-info} associated with the ATLAS paper~\cite{ATLAS:2023oti}).}.
Figure~\ref{fig:13600_500} shows the discovery reach of Run~3 in the $(m_N, V^2_N)$ plane,
assuming the same integrated luminosity as previously, for $\mu=500$, $800$ and $1200~\mathrm{GeV}$.
The increase in sensitivity with respect to Run~2 is clearly visible, extending to active-sterile
neutrino mixings as small as $4\times 10^{-14}$ for $m_N \approx 245~\mathrm{GeV}$
and $\mu \leq 800~\mathrm{GeV}$.

It is interesting to compare the ranges of $m_N$ and $V^2_N$ values that can be probed
by the LHC in the model of Section~\ref{sec:model} with the ones of the standard scenario
in which the sterile neutrino is produced in $W$ decays. As can be seen from Figures~\ref{fig:13600_disc}
and~\ref{fig:13600_500}, in the model considered in this paper, the LHC Run~3 is sensitive to
sterile neutrino masses between a few $10~\mathrm{GeV}$ and about $250~\mathrm{GeV}$,
depending on $\mu$ and $V_N$, and it can probe values of the active-sterile
neutrino mixing in the range
$4 \times 10^{-14} \lesssim V^2_N \lesssim 3 \times 10^{-10}$, depending on $\mu$ and $m_N$.
By contrast, in the standard scenario, the Run~3 of the LHC is not sensitive to sterile neutrinos
heavier than about $20~\mathrm{GeV}$, and its reach is limited to $V^2_N \gtrsim 10^{-8}$
(for the most favourable values of $m_N$) in DV searches~\cite{Drewes:2019fou}.
This is due to the fact that the active-sterile mixing enters the sterile neutrino production
cross section, making very small mixing angles such as the ones suggested by the naive seesaw formula
$V_N \sim \sqrt{m_\nu / m_N}$ inaccessible to the LHC. At the HL-LHC, these numbers improve
to $m_N \lesssim 40~\mathrm{GeV}$ and $V^2_N \gtrsim 5 \times 10^{-10}$~\cite{Drewes:2019fou},
but a large portion of the seesaw parameter space remains out of reach. Indeed, the standard seesaw
mechanism suggests $V^2_N \sim 5 \times 10^{-12}\, (10\, \mathrm{GeV} / m_N)$; values of $V^2_N$
larger than $5 \times 10^{-10}$ require either some fine-tuning in the light neutrino mass matrix,
or a non-standard mechanism such as the inverse seesaw.
In the model of Section~\ref{sec:model}, instead, the LHC can probe even the smallest values of $V^2_N$
that are consistent with neutrino oscillation data, along the black solid curves in the plots of Figures~\ref{fig:13000_500}
and~\ref{fig:13600_500}.

\begin{figure}[t]
    \centering
    \includegraphics[width=0.48\linewidth]{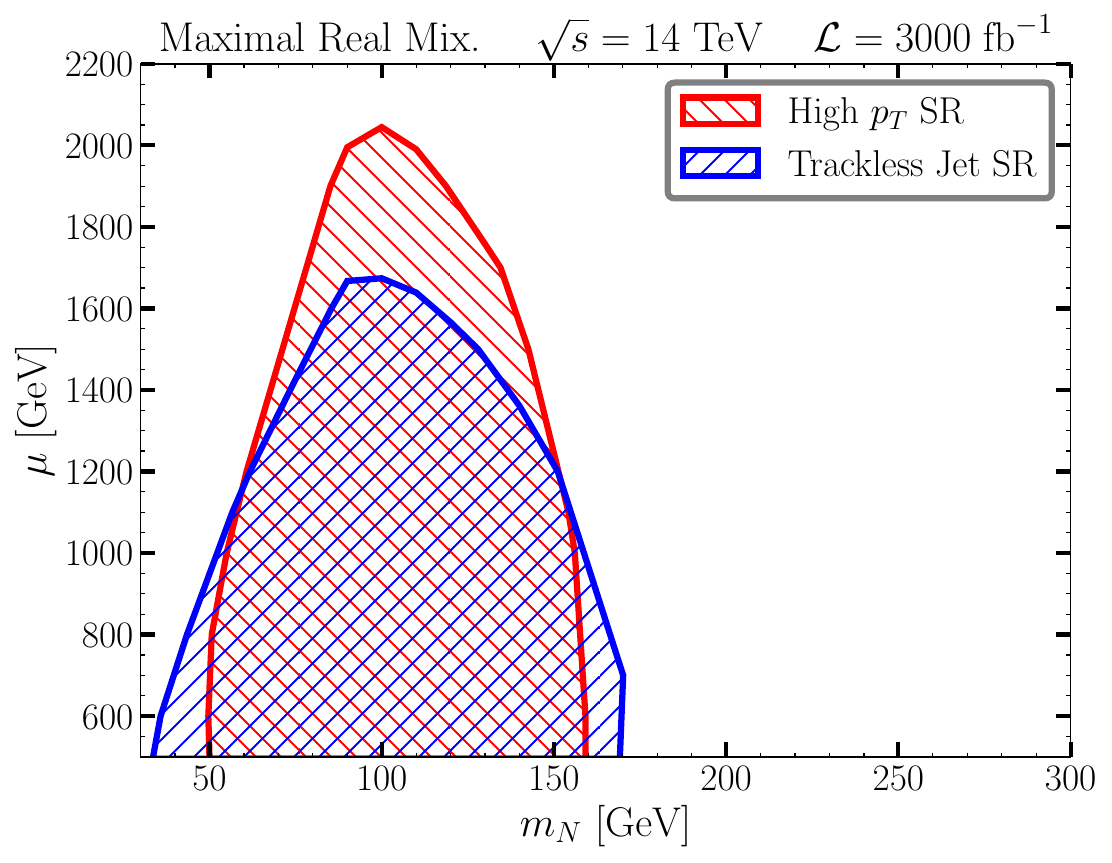}
    \includegraphics[width=0.48\linewidth]{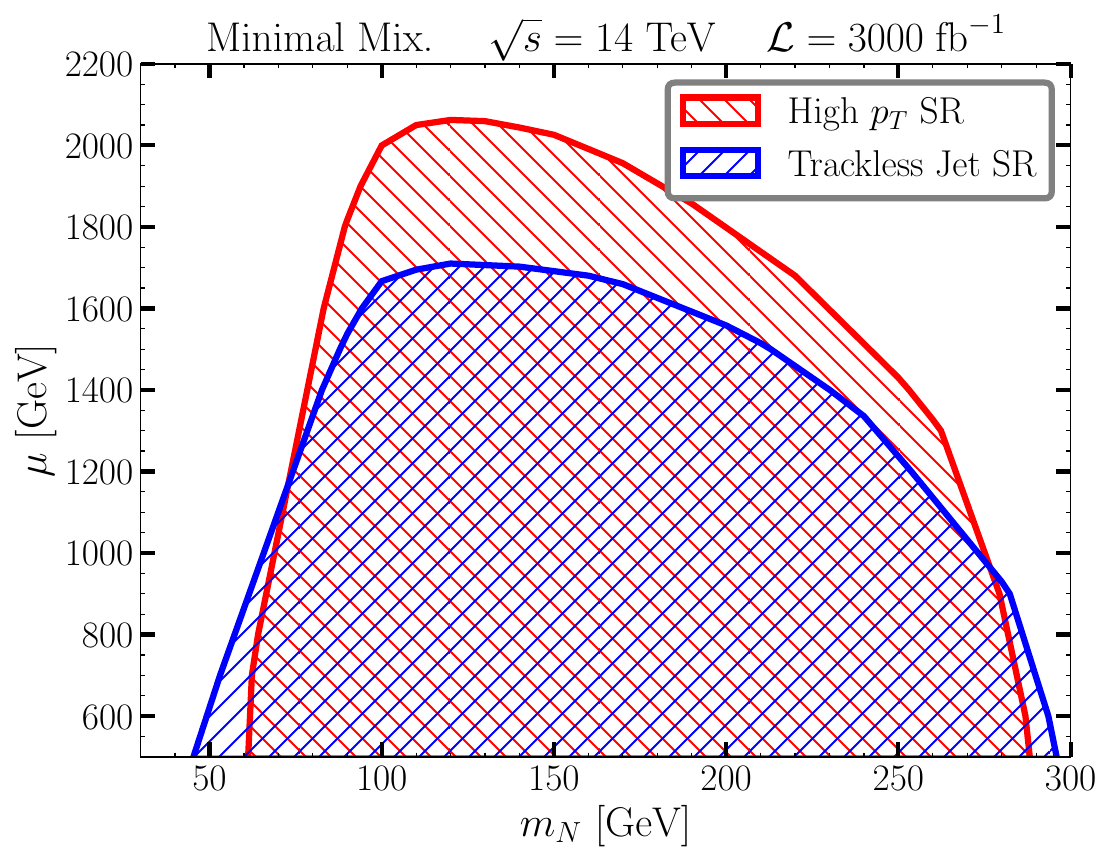}
    \caption{Projected discovery regions in the $(m_N, \mu)$ parameter space of the model of Section~\ref{sec:model}
for the HL-LHC with $\mathcal{L}=3000~\mathrm{fb}^{-1}$, assuming maximal real mixing (left plot) and minimal mixing (right plot).}
    \label{fig:14000_max}
\end{figure}

\begin{figure}[t]
    \centering
    \includegraphics[width=0.48\linewidth]{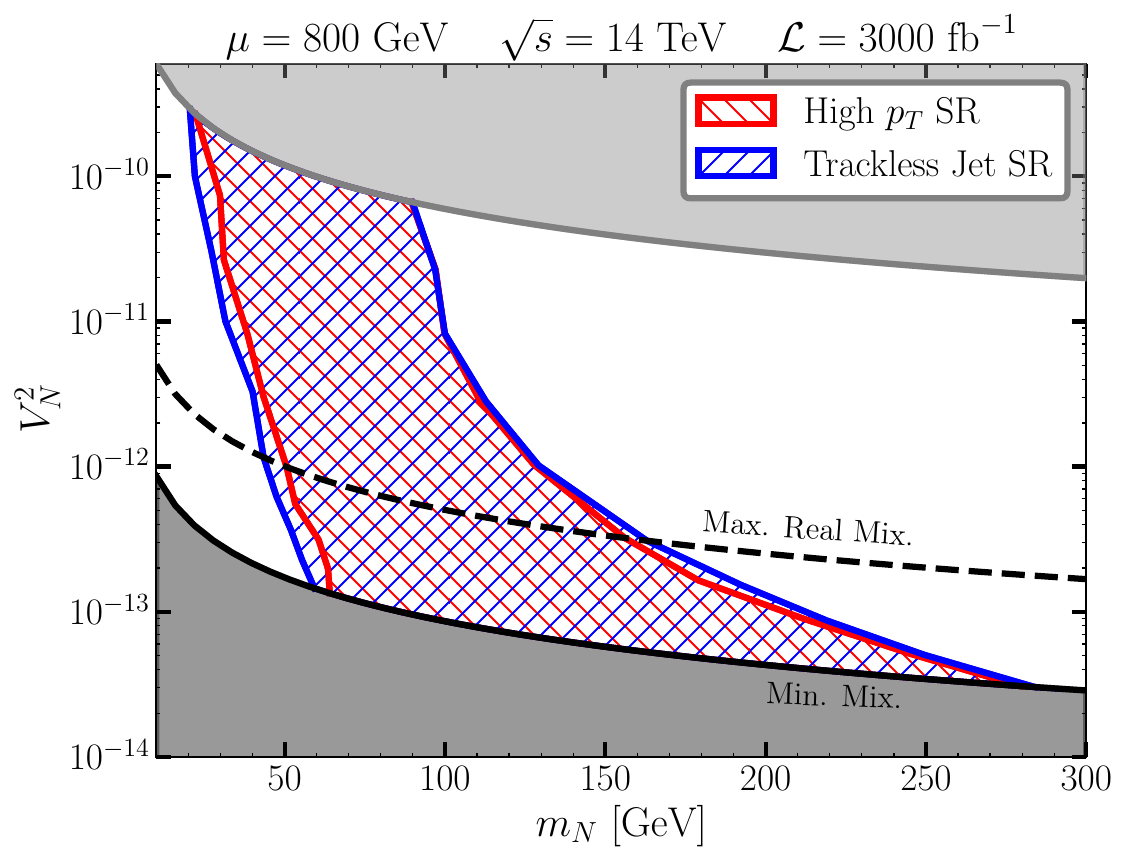}
    \includegraphics[width=0.48\linewidth]{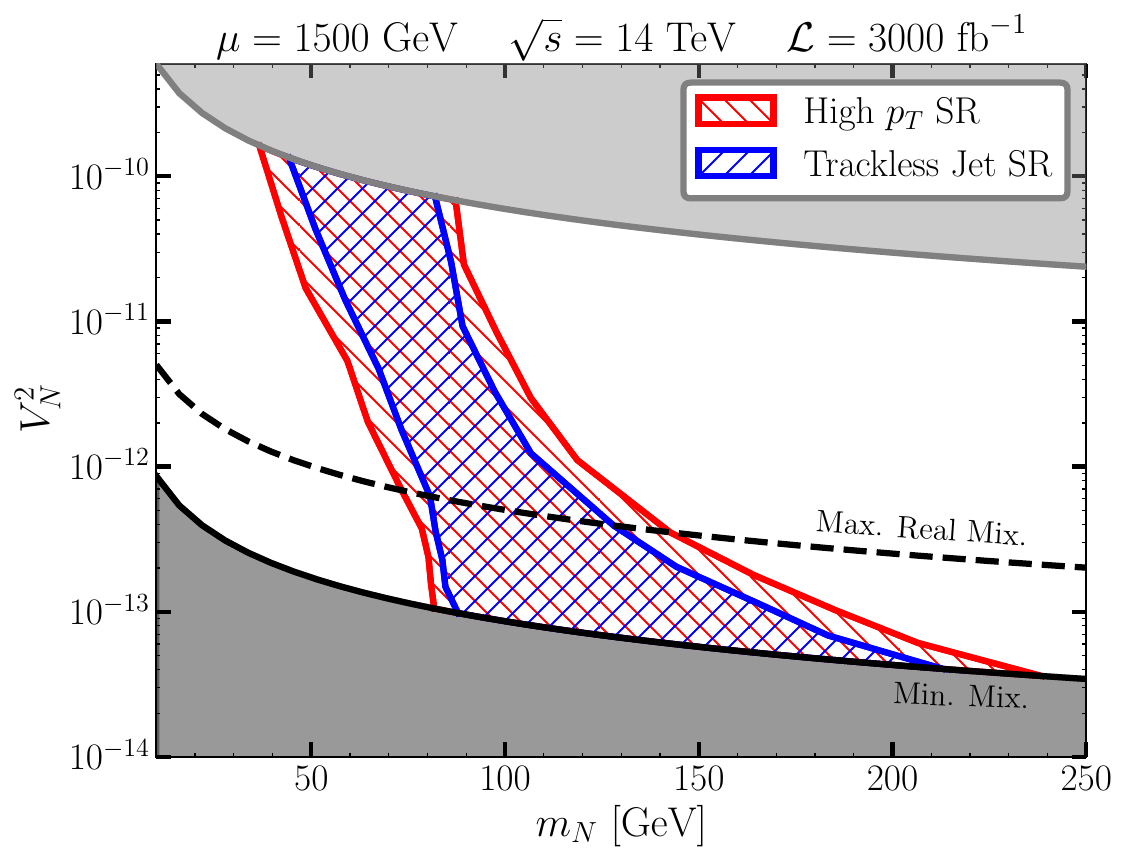}
    \caption{Projected discovery regions in the $(m_N, V^2_N)$ parameter space of the model of Section~\ref{sec:model}
for the HL-LHC with $\mathcal{L}=3000~\mathrm{fb}^{-1}$, assuming $\mu = 800~\mathrm{GeV}$ (left plot)
and $1.5~\mathrm{TeV}$ (right plot).}
    \label{fig:14000_800}
\end{figure}

Let us now consider the high-luminosity LHC, which is expected to collect
$\mathcal{L}=3000~\mathrm{fb}^{-1}$ of $pp$ collision data at a center-of-mass energy $\sqrt{s}=14$~TeV.
Figure~\ref{fig:14000_max} shows the projected discovery areas in the $(m_N, \mu)$ plane
for the HL-LHC with $\mathcal{L}=3000~\mathrm{fb}^{-1}$, assuming maximal real mixing (left plot)
or minimal mixing (right plot).
The increased energy and luminosity with respect to Run~3 greatly enhances the sensitivity
to the higgsino mass parameter, up to $\mu \approx 2~\mathrm{TeV}$ in the High-$p_T$ SR
for both mixing scenarios.
The range of $m_N$ values that can be probed is also wider at the HL-LHC, reaching
$[45, 290]~\mathrm{GeV}$ for small $\mu$ in the minimal mixing case.
The projected discovery areas in the $(m_N, V^2_N)$ plane are displayed in Figure~\ref{fig:14000_800}
for $\mu = 800~\mathrm{GeV}$ (left plot) and $1.5~\mathrm{TeV}$ (right plot).
For $\mu = 800~\mathrm{GeV}$, a large portion of the parameter space is accessible to the HL-LHC,
including sterile neutrino masses as small as $30~\mathrm{GeV}$ for large active-sterile neutrino mixing
(close to the light grey area) and as large as $290~\mathrm{GeV}$ for minimal mixing.
This represents a significant improvement with respect to the expected reach of Run~3.
For larger values of $\mu$, the discovery areas start to shrink and the High-$p_T$ SR provides the best
sensitivity to $m_N$ and $V^2_N$, as can been seen in the right plot of Figure~\ref{fig:14000_800}.

\begin{figure}[t]
    \centering
    \includegraphics[width=0.45\linewidth]{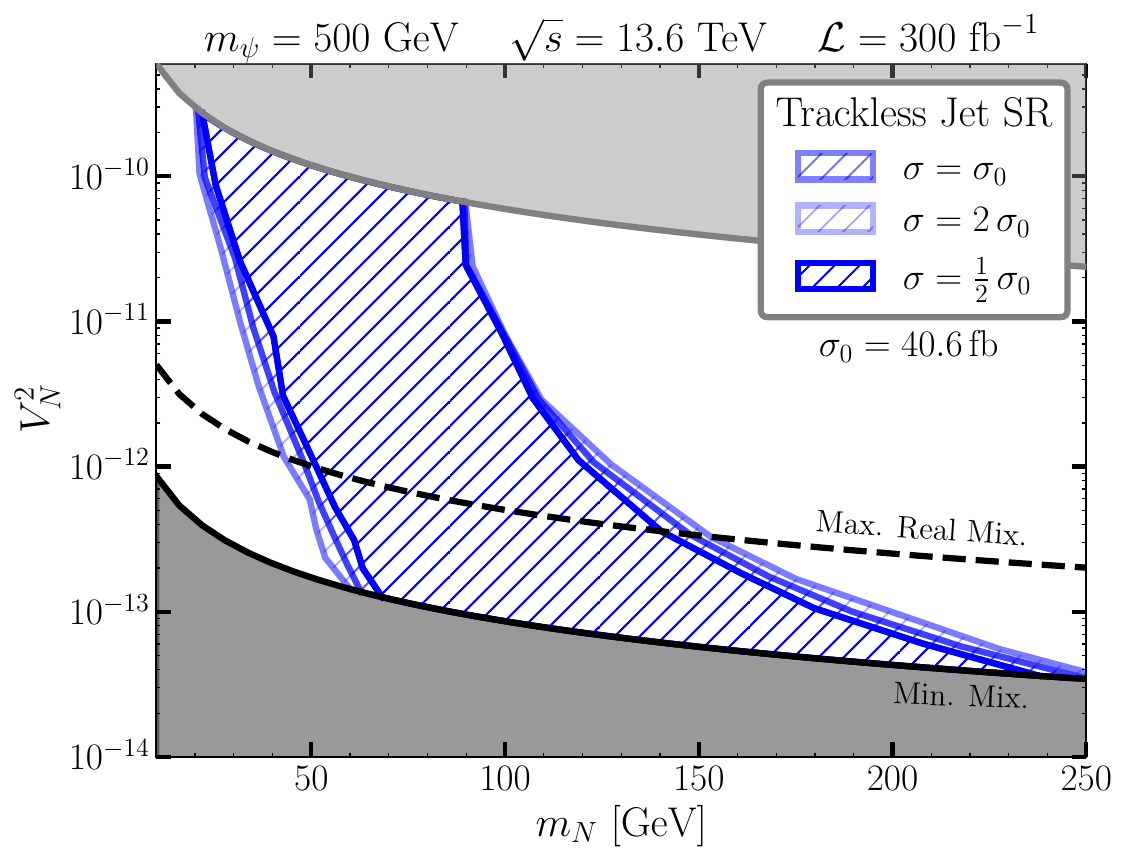}
    \includegraphics[width=0.45\linewidth]{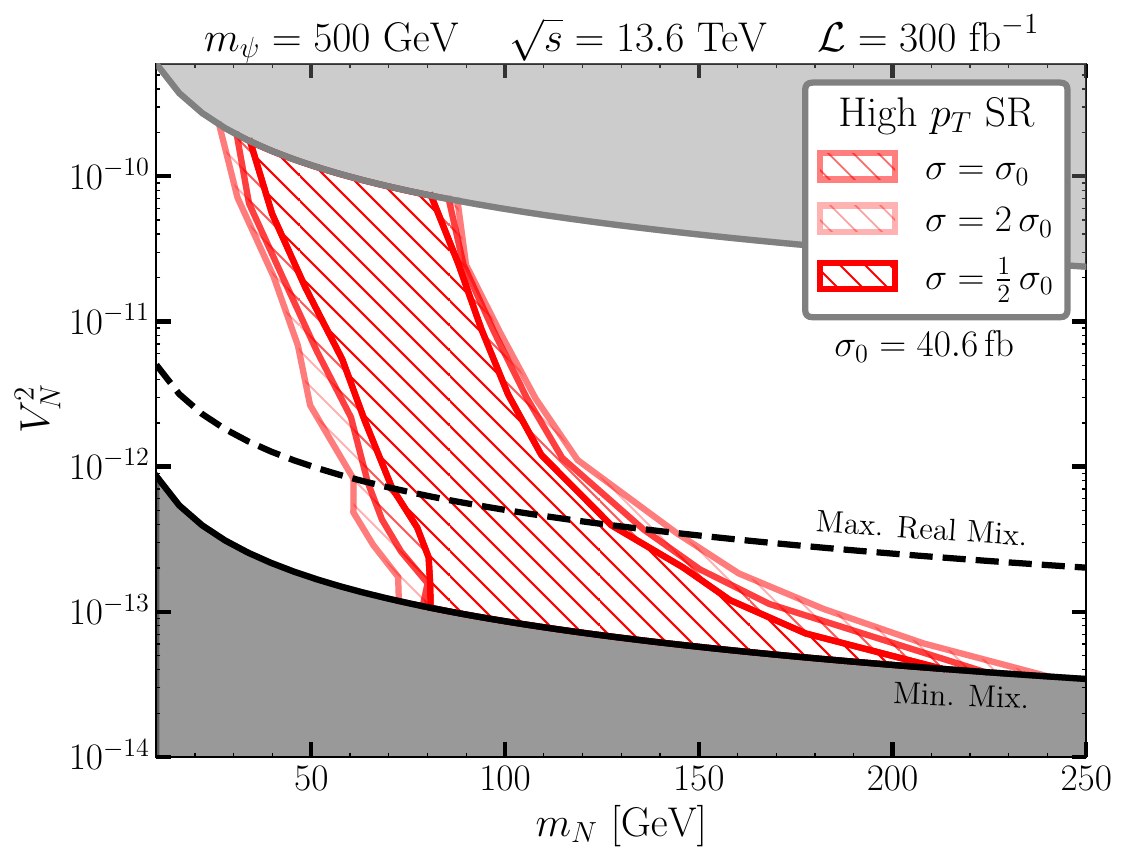} \\ \vspace{7pt}
    \includegraphics[width=0.45\linewidth]{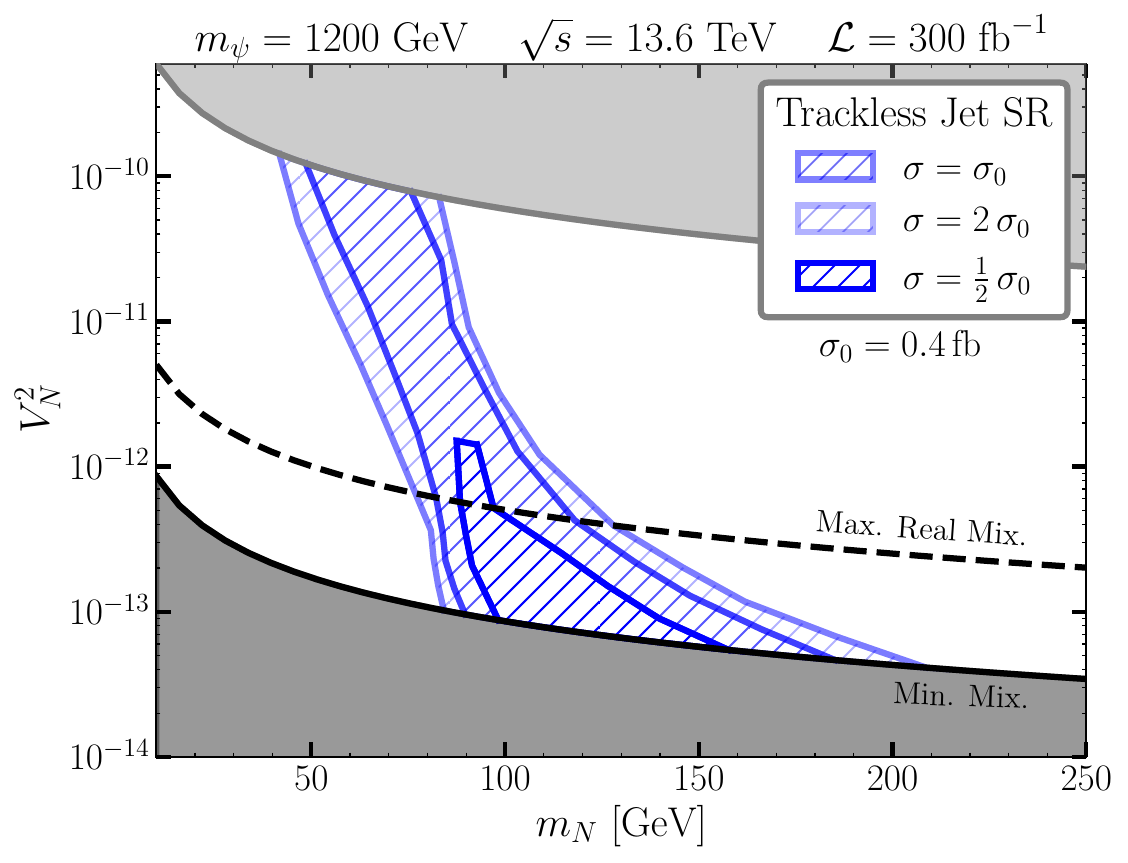}
    \includegraphics[width=0.45\linewidth]{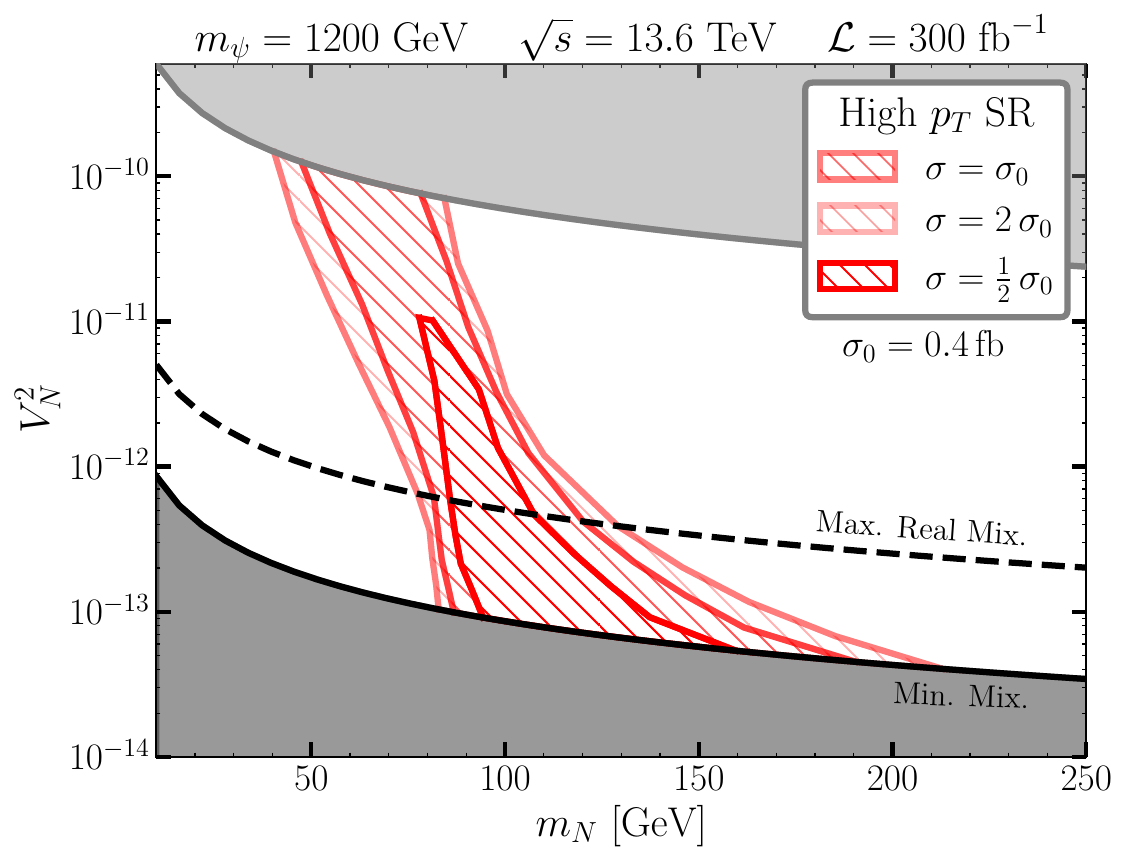}
    \caption{Projected Run~3 discovery areas in the $(m_N, V^2_N)$ plane for the general scenario described in the text,
assuming $m_\psi = 500~\mathrm{GeV}$ in the upper panels and $m_\psi = 1.2~\mathrm{TeV}$ in the lower panels.
Each plot displays the discovery regions for three different values of the $pp \to \psi \psi$ cross section:
$\sigma = \sigma_0$, $\sigma = 2 \sigma_0$ and $\sigma = \sigma_0 / 2$,
where $\sigma_0$ is the higgsino pair production cross section evaluated at $\mu = m_\psi$.
The results are presented separately for the Trackless jet (left panels) and High-$p_T$ (right panels) signal regions.}
    \label{fig:13600_psi}
\end{figure}

While the results presented above were derived within the model of Section~\ref{sec:model},
they can be generalized to some extent to a broader class of models in which the sterile neutrino is produced
in the decays of heavier particles $\psi$ that are themselves pair-produced in proton-proton collisions
with an electroweak-size cross section. Let us more specifically consider the scenario in which $\psi$ decays
promptly to $N + V$ ($V = W, Z$) with an $\approx 100\%$ branching ratio.
On general grounds we do not expect a single particle $\psi$, but several charged and neutral states
originating from the same $SU(2)_L$ multiplet. We therefore consider both decays
$\psi^0 \to N + Z$ and $\psi^\pm \to N + W^\pm$.
As far as sterile neutrino production and decay are concerned, this scenario and the model of Section~\ref{sec:model}
only differ in the following three points:
{\it (i)} the $pp \to \psi \psi$ production cross section $\sigma (m_\psi)$ is in general
different from the higgsino pair production cross section
$\sigma_0 (\mu) \equiv \sigma (\tilde{\chi}^{\pm}_1 \tilde{\chi}^0_1) + \sigma (\tilde{\chi}^{\pm}_1 \tilde{\chi}^0_2)
+ \sigma (\tilde{\chi}^+_1 \tilde{\chi}^-_1) + \sigma (\tilde{\chi}^0_1  \tilde{\chi}^0_2)$;
{\it (ii)} the relative size of the production channels $p p \to Z Z N N$, $p p \to W^\pm Z N N$
and $p p \to W^+ W^- N N$;
{\it (iii)} the relative size of the active-sterile mixing angles $V_{N e}$, $V_{N \mu}$ and $V_{N \tau}$.
Since the recasted ATLAS search targets multiple energetic jets, and the $W$ and $Z$ bosons
have very close branching fractions into hadrons (a statement that remains true when hadronic tau
decays are taken into account), we can ignore point {\it (ii)} to a good approximation.
We also disregard point {\it (iii)}, with the caveat that a larger $V_{N \tau}$ results in a larger
proportion of jets in $N$ decays, due to hadronic $\tau$'s.
In order to quantify the impact of point {\it (i)} on our results, 
we show in Figure~\ref{fig:13600_psi} the projected Run~3 discovery areas in the $(m_N, V^2_N)$ plane
for three different values of the $pp \to \psi \psi$ cross section:
$\sigma (m_\psi) = \sigma_0 (m_\psi)$, $2 \sigma_0 (m_\psi)$ and $\sigma_0 (m_\psi) / 2$,
where $\sigma_0 (m_\psi)$ is the higgsino pair production cross section evaluated at $\mu = m_\psi$.
Two values of the heavy particle mass are considered ($m_\psi = 500~\mathrm{GeV}$ in the upper panels,
$m_\psi = 1.2~\mathrm{TeV}$ in the lower panels), and the discovery regions corresponding to the Trackless jet
and High-$p_T$ signal regions are presented in different panels.
One can see that for large enough production cross sections (upper plots with $m_\psi = 500~\mathrm{GeV}$,
corresponding to $\sigma_0 = 40.6\, \mathrm{fb}$), the discovery areas are not significantly affected
by variations of $\sigma (m_\psi)$ around $\sigma_0$ (with however some more notable differences
in the High-$p_T$ SR), while they are very sensitive to the actual value
of $\sigma (m_\psi)$ for small production cross sections (lower plots with $m_\psi = 1.2~\mathrm{TeV}$,
corresponding to $\sigma_0 = 0.4\, \mathrm{fb}$). In particular, for $m_\psi = 1.2~\mathrm{TeV}$
and $\sigma (m_\psi) = \sigma_0 (m_\psi) / 2$, Run~3 will not be able to probe values of $V^2_N$
above $1.4 \times 10^{-12}$ in the Trackless jet SR ($10^{-11}$ in the High-$p_T$ SR),
while values up to $\approx 10^{-10}$ are accessible for $\sigma (m_\psi) = \sigma_0 (m_\psi)$ in both SRs.

%%%%%%%%%%%%
\section{Conclusions}
\label{sec:conclusion}
%%%%%%%%%%%%

In this work, we have studied the sensitivity of displaced vertex searches at the LHC to heavy sterile neutrinos
that are pair-produced with an electroweak-size cross section in proton-proton collisions.
As a case study, we considered the $R$-parity violating supersymmetric model of Ref.~\cite{Lavignac:2020yld},
in which the sterile neutrino is the supersymmetric partner of a pseudo-Nambu-Goldstone boson
and is involved in the generation of light neutrino masses.
In this model, the sterile neutrinos are produced along with an electroweak gauge boson in higgsino decays.
They subsequently decay via their mixing with active neutrinos, giving rise to observable displaced vertices.

We first obtained the constraints set by the LHC Run~2 on this model by recasting the ATLAS search
for displaced vertices and multiple jets of Ref.~\cite{ATLAS:2023oti}.
These constraints are presented as $95\%$ C.L. exclusion regions in the $(m_N, V^2_N)$ plane
for fixed values of the higgsino mass parameter $\mu$ in Figure~\ref{fig:13000_500}.
For $\mu \leq 800~\mathrm{GeV}$, sterile neutrino masses ranging from $30~\mathrm{GeV}$
to $215~\mathrm{GeV}$ are excluded, depending on the active-sterile mixing $V_N$.
Conversely, $V^2_N$ values between $4 \times 10^{-14}$ and $2 \times 10^{-10}$
(where the upper bound comes from fine-tuning considerations) are excluded, depending on $m_N$.
Then, extrapolating the procedure to higher energies
and luminosities, we assessed the discovery reach of Run~3 and of the high-luminosity LHC.
As can be seen from Figures~\ref{fig:14000_max} and~\ref{fig:14000_800},
the HL-LHC will be able to probe sterile neutrino masses between about $20~\mathrm{GeV}$ and $295~\mathrm{GeV}$,
depending on $V_N$ and $\mu$,
and values of $V^2_N$ ranging from $3 \times 10^{-14}$ to $3 \times 10^{-10}$.
By contrast, in the standard scenario where the sterile neutrino is produced through its
mixing with active neutrinos, the HL-LHC reach is limited to $m_N \lesssim 40~\mathrm{GeV}$
and $V^2_N \gtrsim 5 \times 10^{-10}$ in displaced vertex searches~\cite{Drewes:2019fou}.
Coming back to the model of Section~\ref{sec:model}, it is interesting to note that for $m_N$
around $100~\mathrm{GeV}$, higgsino masses up to $\mu \approx 2~\mathrm{TeV}$ will be accessible
at the HL-LHC.

We have investigated to which extent these conclusions
can be generalized to a broader class of models in which the sterile neutrino is produced
along with a $W$ or $Z$ boson in the decays of heavier particles $\psi$, which are themselves pair-produced
with an electroweak-size cross section. The main difference between this scenario
and the model of Section~\ref{sec:model} lies in the dependence of the pair production cross section
$\sigma$ on the heavy particle mass $m_\psi$.
For large enough production cross sections (of order a few tens of fb),
the LHC discovery reach is not significantly affected by variations of $\sigma (m_\psi)$
around $\sigma_0 (m_\psi)$, where $\sigma_0 (m_\psi)$ is the higgsino pair production cross section
for $\mu = m_\psi$. By contrast, for cross sections below $1~\mathrm{fb}$,
the LHC reach is very sensitive to the actual value of $\sigma (m_\psi)$.
For such small production cross sections, the expected HL-LHC sensitivity to $m_N$ and $V_N$
is significantly reduced with respect to the model of Section~\ref{sec:model}
if $\sigma (m_\psi) \lesssim \sigma_0 (m_\psi) / 2$.

\section*{Acknowledgements}

We thank Giovanna Cottin for useful discussions.
The work of S.L. is supported in part by the European Union's Horizon Europe research and innovation programme
under the Marie Sklodowska-Curie Staff Exchange grant agreement No. 101086085 -- ASYMMETRY.
The research of A.M. was supported in part by Perimeter Institute for Theoretical Physics.
Research at Perimeter Institute is supported by the Government of Canada through the
Department of Innovation, Science, and Economic Development, and by the Province
of Ontario through the Ministry of Colleges and Universities.
S.T. acknowledges the hospitality and support of the Jo\v{z}ef Stefan Institute and of the Institut de Physique Th\'eorique during the research stays undertaken in the course of this work.

\appendix

%%%%%%%%%%%%%%%
\section{Validation of the recasting procedure}
\label{sec:validation}
%%%%%%%%%%%%%%%

\begin{figure}[t]
    \centering
    \includegraphics[width=0.7\linewidth]{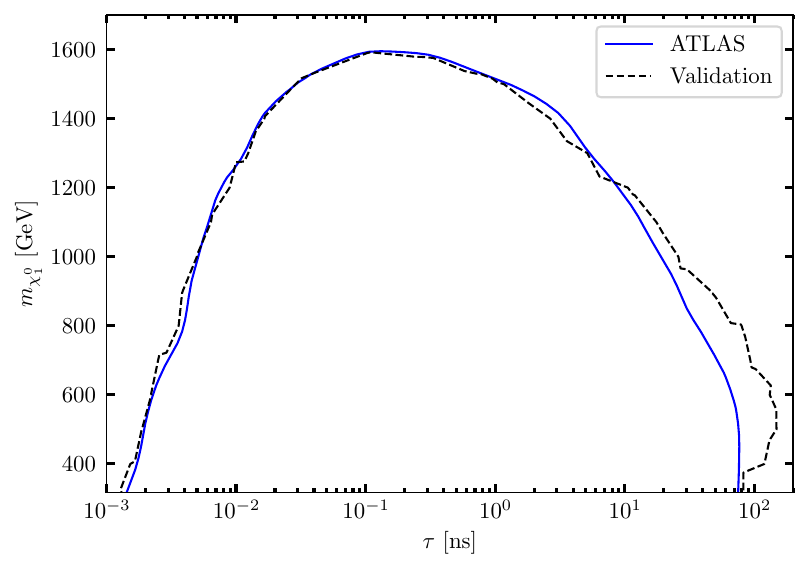}
    \caption{Comparison between ATLAS observed exclusion limit and our validation of the recasting procedure, for the electroweak RPV model.}
    \label{fig:validation}
\end{figure}

We validate our implementation of the recasting procedure by reobtaining the observed exclusion limit presented in the ATLAS search~\cite{ATLAS:2023oti} for the electroweak RPV model.
In this case, the LLPs with lifetime $\tau$ are the lightest chargino $\chi_1^\pm$ and the two lightest neutralinos $\chi_{1,2}^0$, which are pure-higgsino states and degenerate in mass, while all other SUSY particles are decoupled.
We generate $pp\to \chi_1^0\,\chi^0_2\,,\chi_1^0\,\chi_1^\pm\,,\chi_2^0\,\chi_1^\pm\,,\chi_1^+\chi_1^-$ with up to one extra jet.
We scan the $(\tau,m_{\chi^0_1})$ parameter space and apply our implementation of the reinterpretation material to obtain acceptance $\times$ efficiency values.
The signal counts are obtained by rescaling to the appropriate cross section, obtained with \texttt{Resummino 3.1.2} at NLO+NLL precision, and luminosity $\mathcal{L}=139\,\mathrm{fb}^{-1}$.
We implement the two-bin statistical model in \texttt{pyhf 0.7.6}~\cite{pyhf, pyhf_joss}
combining the Trackless jet and High-$p_T$ signal regions, as described in~\cite{ATLAS:2023oti}.
Observed counts and background estimations, along with their uncertainties, are reported in the ATLAS search.
Signal uncertainties involve propagating, for each point in the parameter space, different sources such as uncertainties in the reconstruction efficiencies, Monte Carlo acceptances, cross section and luminosity.
Providing an accurate estimate of the signal uncertainty for this model is beyond the scope of the validation task, and instead we apply a flat $10\%$ relative uncertainty for all the points.
Varying this value between $0\%-20\%$ does not change our results.
The resulting exclusion curve and the exclusion curve provided by the ATLAS search are shown in Figure~\ref{fig:validation}.
Our implementation matches the ATLAS contour below $\tau = 10\;\mathrm{ns}$, while the excess seen in the large lifetime regime in consistent with other independent validation procedures~\cite{code-LLP-recasting}.

\newpage

\bibliographystyle{bibstyle}
\bibliography{sample.bib}

\end{document}